\documentclass[11pt]{article}
\usepackage{mathrsfs}
\usepackage{natbib,graphicx,setspace,lscape,longtable}
\usepackage{mathrsfs,amsmath,amsthm,amssymb,color}
\usepackage{natbib,epsfig,graphicx,pdfpages}
\usepackage{rotating}
\usepackage{subfigure}
\usepackage{graphicx}
\usepackage{enumerate}
 \usepackage{hyperref}
 \usepackage{tikz}
\usepackage{fixltx2e}
\usetikzlibrary{shapes,arrows}
\bibpunct{(}{)}{;}{a}{,}{,}

\newtheorem{theorem}{Theorem}
\newtheorem{lemma}{Lemma}
\newtheorem{proposition}{Proposition}
\newtheorem{remark}{Remark}
\usepackage{amsthm}

\newtheorem{assumption}{{Assumption}}
\def\beq{\begin{equation}}
\def\eeq{\end{equation}}
\def\beqr{\begin{eqnarray}}
\def\eeqr{\end{eqnarray}}
\def\beqrs{\begin{eqnarray*}}
\def\eeqrs{\end{eqnarray*}}
\def\bet{\begin{theorem}}
\def\eet{\end{theorem}}
\def\bel{\begin{lemma}}
\def\eel{\end{lemma}}
\def\bep{\begin{proposition}}
\def\eep{\end{proposition}}
\def\bg{\begin{figure}[tbph]\begin{center}}
\def\eg{\end{center}\end{figure}}

\def\bc{\begin{center}}
\def\ec{\end{center}}

\def\wt{\widetilde}
\def\wh{\widehat}

\def\diag{\mbox{diag}}

\numberwithin{equation}{section}

\newcommand{\Cov}{\textnormal{Cov}}

\newcommand{\bA}{{\mathbf A}}
\newcommand{\bB}{{\mathbf B}}

\newcommand{\bG}{{\mathbf G}}
\newcommand{\bH}{{\mathbf H}}
\newcommand{\bI}{{\mathbf I}}

\newcommand{\bL}{{\mathbf L}}
\newcommand{\bM}{{\mathbf M}}

\newcommand{\bQ}{{\mathbf Q}}
\newcommand{\bP}{{\mathbf P}}
\newcommand{\bR}{{\mathbf R}}
\newcommand{\bS}{{\mathbf S}}

\newcommand{\bV}{{\mathbf V}}

\newcommand{\ba}{{\mathbf a}}
\newcommand{\bb}{{\mathbf b}}

 \newcommand{\bfc}{{\mathbf c}}
\newcommand{\be}{{\mathbf e}}
\newcommand{\bff}{{\mathbf f}}
 
\newcommand{\bh}{{\mathbf h}}

\newcommand{\br}{{\mathbf r}}

\newcommand{\bu}{{\mathbf u}}

\newcommand{\bw}{{\mathbf w}}
\newcommand{\bx}{{\mathbf x}}
\newcommand{\by}{{\mathbf y}}
\newcommand{\bz}{{\mathbf z}}

\newcommand{\bfeta}  {\boldsymbol{\eta}}

\newcommand{\bSigma}{\boldsymbol{\Sigma}}

\newcommand{\bve}{\mbox{\boldmath$\varepsilon$}}

\newcommand{\bTheta} {\boldsymbol{\Theta}}
\newcommand{\bPhi} {\boldsymbol{\Phi}}

\newcommand{\bGamma} {\boldsymbol{\Gamma}}

\newcommand{\bD}{{\mathbf D}}
\newcommand{\bzero}{{\mathbf 0}}

\newcommand{\ve}{{\varepsilon}}
\renewcommand{\epsilon}{{\ve}}
\renewcommand{\hat}{\widehat}

\def\wt{\widetilde}

\newcommand{\tr}{\mbox{tr}}
\def\JRSSB{{\sl Journal of the Royal Statistical Society}, {\bf B}}

\def\BKA{{\sl Biometrika}}
\def\JASA{{\sl Journal of the American Statistical Association}}

\begin{document}

\title{\bf Modeling High-Dimensional Time Series: A 
Factor Model with Dynamically Dependent Factors and Diverging Eigenvalues}

\author{
Zhaoxing Gao$^1$ and Ruey S. Tsay$^2$ \\
$^1$Department of Mathematics, Lehigh University\\
$^2$Booth School of Business, University of Chicago
}


\maketitle

\begin{abstract}
This article proposes a new approach to modeling high-dimensional time series by 
treating a $p$-dimensional time series as a nonsingular linear transformation of certain 
common factors and idiosyncratic components.
Unlike the approximate factor models, we assume that 
the factors capture all the non-trivial dynamics of the data, but the cross-sectional dependence may be explained by both the factors and the idiosyncratic components. Under the proposed model, (a) the factor process is dynamically dependent and the idiosyncratic component is a white noise process, and  (b) the largest eigenvalues 
of the covariance matrix of the idiosyncratic 
components  may diverge to infinity as the dimension $p$ increases. 
We propose a white noise testing procedure for high-dimensional time series 
to determine the number of white noise components and, hence, the number of 
common factors, and  introduce a projected Principal 
Component Analysis (PCA) to eliminate the diverging effect of the idiosyncratic noises. 
Asymptotic properties of the proposed method are established for both fixed $p$ and 
diverging $p$ as the sample size $n$ increases to infinity. 
We use both simulated data and real examples to assess the performance 
of the proposed method. 
We also compare our method with two commonly used methods in the literature 
concerning the forecastability of the extracted factors 
and find that the proposed approach not only provides interpretable results, 
but also performs well in out-of-sample forecasting. Supplementary materials of 
the article are available online.

\end{abstract}


\newpage

\section{Introduction}
Advances in information technology make large data sets widely accessible nowadays. In many 
applications, the data consist naturally of  high-dimensional time series.  
For example, the returns of a large number of assets form a high-dimensional time series 
and play an important role in asset pricing, portfolio allocation, and risk management. 
Large panel time series data are commonplace in economics and biological studies.  Environmental studies often employ high-dimensional time series consisting of a large number of pollution indexes 
collected at many monitoring stations and  over periods of time. 
However, modeling high-dimensional time series   
is always challenging because the commonly used Vector Autoregressive (VAR) or Vector-Autoregressive Moving-Average (VARMA) models are not practically applicable when the dimension is high. In particular, unregularized VARMA models often suffer the difficulties 
of over-parameterization and lack of identifiability as discussed in \cite{TiaoTsay_1989}, \cite{lutkepohl2006}, and \cite{Tsay_2014}. 
 Therefore, dimension reduction or structural specification becomes a necessity 
 in applications of high-dimensional time series. Indeed, 
various methods have been developed in the literature for multivariate 
time series analysis, including the scalar component models of \cite{TiaoTsay_1989}, the LASSO regularization of VAR models by \cite{ShojaieMichailidis_2010} and \cite{SongBickel_2011}, the sparse VAR model based on partial spectral coherence by \cite{Davis2012}, and the factor modeling in \cite{StockWatson_2005}, \cite{BaiNg_Econometrica_2002}, \cite{forni2005}, \cite{LamYaoBathia_Biometrika_2011},  and \cite{lamyao2012}, among others. However, the complexity of the dynamical dependence in high-dimensional time series requires further investigation, as extracting dynamic information from the data plays an important role  in modeling and forecasting  serially dependent data.

This article marks a further development in  factor modeling of high-dimensional 
time series.  Factor models are commonly used in finance, economics, and statistics. 
For example, asset returns are often modeled as functions of a small number of factors, see \cite{StockWatson_1989} and \cite{StockWatson_1998}. 
Macroeconomic variables of multiple countries are often found to have common movements, see \cite{GregoryHead1999} and \cite{forni2000a}. From the statistical perspective, a modeling approach that can reveal the common structure of the 
series and provide accurate estimation of a specified model is highly valuable in understanding 
the dynamic relationships of the data. To this end, we first briefly introduce the traditional approximate factor models and some existing estimation procedures under different assumptions in the literature. Let $\by_t=(y_{1t},...,y_{pt})'$ be a $p$-dimensional zero-mean time series. 
The approximate factor model for $\by_t$ assumes the form 
\begin{equation}\label{tra-f}
\by_t=\bA\bx_t+\bve_t,
\end{equation}
where $\bx_t$ is a $r$-dimensional latent factor process, $\bA\in R^{p\times r}$ is an associated (full-rank) factor loading matrix,  $\bve_t$ is the idiosyncratic component, and 
$\bx_t$ and $\bve_t$ are independent. To the best of our knowledge, there are at least three main statistical procedures to estimate the common factors and the associated loading matrix under various assumptions on the factor and idiosyncratic terms.
The first procedure is based on the principal component analysis (PCA), see \cite{BaiNg_Econometrica_2002}, \cite{Bai_Econometrica_2003}, \cite{fan2013} with pervasive factors, and \cite{onatski2012} with weakly influential factors, and the references therein. 
The second procedure is based on the eigen-analysis of the auto-covariance matrices, see \cite{LamYaoBathia_Biometrika_2011} and \cite{lamyao2012}, among others. These authors assume that all components of $\bx_t$ are dynamically dependent, reflecting the nature of time series data, and $\bve_t$ is a vector white noise process, which has no serial correlations. Although the model considered in \cite{LamYaoBathia_Biometrika_2011} and \cite{lamyao2012} forms a subclass of 
those in \cite{BaiNg_Econometrica_2002}, the recovered factors are different from those in the latter because they are all dynamically dependent. 
The third approach is slightly different from the previous two and allows $\by_t$ to depend also on factors 
involving lagged variables, which can also be written as Equation (\ref{tra-f}) with $\bx_t$ 
consisting of all dynamic factors. See \cite{forni2000b}, \cite{forni2005}, \cite{forni2015}, and the references therein. This third approach is based on the eigenvalues and principal components of spectral density matrices and, hence, is  a 
frequency-domain analysis. 
But some recent studies extend the approach to time-domain analysis; see, for instance, 
\cite{HallinLippi2013} and \cite{PenaYohai2016}. 
For more details on the differences between the first and the third approaches, see \cite{fan2013}.

In this paper, we study the common dynamic components in high-dimensional, and possibly highly cross-sectionally correlated, time series data. We propose a new factor model under which the observed high-dimensional time series 
$\by_t$ is a nonsingular linear transformation of a $r$-dimensional 
common factor process, which is dynamically dependent,  
and a $(p-r)$-dimensional idiosyncratic component, which is a white noise series. 
In other words, we assume the factors capture all the non-trivial dynamics of the data, but the cross-sectional dependence may be explained by both the factors and the idiosyncratic components. 
This is different from the approximate factor model, where factors capture most of the cross-sectional dependence, while the idiosyncratic terms may contain some non-trivial temporal dependence.   To a large degree and in extracting the common dynamic components of the data, our approach is closer to the dynamic factor models of \cite{forni2000a,forni2000b,forni2005}, even though its form is static according to, for example, \cite{boivin-ng2005}. In fact, 
the proposed new factor model is in line with that of \cite{TiaoTsay_1989} 
and \cite{gaotsay2018}, and 
assumes that the idiosyncratic component is white noise in the sense that $\by_t$ has $(p-r)$ scalar components of order $(0,0)$. See  \cite{TiaoTsay_1989} and Section 2 for details. 
However, our proposed modeling approach is different from 
those of the aforementioned two papers, because 
\cite{TiaoTsay_1989} assumes $p$ is fixed and 
\cite{gaotsay2018} considers $p = o(n^{1/2})$,  where $n$ is the sample size, so that 
they can employ the canonical correlation analysis (CCA). In this paper, 
we do not employ CCA and, hence, can relax the constraints on $p$ as $n$ increases.

Similar to that of \cite{LamYaoBathia_Biometrika_2011} and \cite{lamyao2012}, 
we first apply the eigen-analysis on certain auto-covariance matrices to obtain the loading matrix 
associated with the common factors. But we propose a projected PCA method to estimate 
the loading matrix associated with the idiosyncratic component in the presence of diverging noise effect. Our estimator is constructed by first projecting $\by_t$ onto the directions which capture little dynamic dependencies, and the resulting projected coordinates would essentially be a vector white noise. We then perform the PCA between the original data $\by_t$ and the projected white noise coordinates, and the eigenspace associated with the small eigenvalues would be able to mitigate the diverging effect of the white noise idiosyncratic components;  see Section 2 for details. 
In addition, we propose a new method to estimate the common factors so that the resulting 
estimated common factors are not affected by the idiosyncratic component $\bve_t$. 
Specifically, in the presence of diverging noise components, we project the observed data into the orthogonal direction of those diverging noise components  to mitigate the effect of the idiosyncratic 
component in estimating the common factors. 
Furthermore, to overcome the difficulties associated with the behavior of eigenvalues of 
a large random matrix, we consider a white noise testing procedure to determine the number 
of common factors. This testing procedure is found to be more reliable than the the ratio-based method currently used in the literature to extract the dynamically dependent factors. Details of the testing  procedure are given in Section 2.3.  

The idea of using white noise test to determine the number of factors has been used in \cite{panyao2008} when the dimension is relatively small. However, their approach needs to solve a constrained optimization problem step-by-step and 
it cannot be extended to 
the high-dimensional case directly. When the factors in (\ref{tra-f}) have different levels of strength, \cite{lamyao2012} proposed a two-step estimation procedure to successively identify two groups of factors with top two strengths. Under an independent structure of the noises, \cite{lietal2017} provided an exact description of the phase transition phenomenon whether a factor is strong enough to be detected with the observed sample singular values and proposed a new ratio-based estimator  for determining the number of factors, which is shown to be  more robust against possibly multiple levels of factor strengths. However, they all rely on relatively pervasive factors in the sense that their strength is (much) stronger than that of the idiosyncratic components, which is different from the setting considered in this paper. The proposed approach still works when the factors have multiple levels of factor strengths and the theory can readily be modified accordingly. 
For simplicity, we only consider factors with the same strength in this paper. 

We conduct simulation studies to assess the performance of the proposed modeling procedure in 
finite samples and to compare it with the method in \cite{LamYaoBathia_Biometrika_2011}. 
When the largest eigenvalues of the covariance matrix of the idiosyncratic component are 
diverging, the results 
show that  the method of \cite{LamYaoBathia_Biometrika_2011} may encounter 
estimation errors. 
We further apply the proposed method to two real examples, and the numerical results 
suggest that the factors recovered by our approach not only have reasonable interpretations but also fare well in  prediction. On the other hand, even under the setting of \cite{LamYaoBathia_Biometrika_2011}, the ratio-based method could fail if the covariance of the noises is large. See the simulation results in the supplementary materials. Furthermore, as the factors play an important role in diffusion index models as in \cite{StockWatson_2002a} and \cite{StockWatson_2002b}, for  the purpose of out-of-sample forecasting, we also  compared the predictive ability of the factors extracted by different methods in an empirical example. The results suggest that the dynamically dependent factors extracted by our method fare better in predicting the U.S.  Consumer Price Indexes. 

The contributions of this paper are multi-fold. First,  the proposed model is flexible in 
application. 
It allows a variety of structures for the common factors and the idiosyncratic components. 
Second, the proposed estimation method can eliminate the effect of the idiosyncratic term 
in estimating the common factors. This is achieved by using the projected PCA method 
if the dimension $p$ is low.  When the dimension is high, we assume that a few largest 
eigenvalues of the covariance matrix of the idiosyncratic noise term are diverging, 
which is a reasonable assumption in the high-dimensional setting and the diverging eigenvalues can explain some of the cross-sectional dependence. The projected PCA then helps to 
mitigate the effect of the diverging part of the noise covariance matrix. 
Third, we propose a procedure based on a white noise test for multivariate (and high-dimensional) time series to determine the number of common factors $r$. Under the assumption that the idiosyncratic term is a vector white noise, the limiting distribution of the test statistic used is available in closed form. This testing procedure is shown to be more reliable than the ratio-based 
method available in the literature in the sense that our approach can extract the dynamically dependent factors more accurately.

The rest of the paper is organized as follows. We introduce the proposed model, 
estimation methodology, and modeling procedure in Section 2, 
including a flowchat for the entire modeling procedure. 
 In Section 3, we study the theoretical properties of the proposed model and its associated 
 estimates. 
Section 4 illustrates the performance of the proposed model using both 
simulated and real data sets. Section 5 provides some discussions and concluding remarks. 
All technical proofs and an additional real example are relegated to an online supplement. Throughout the article,
 we use the following notation.  $||\bu||_2 = (\sum_{i=1}^{p} u_i^2)^{1/2} $
is the Euclidean norm of a $p$-dimensional vector  
$\bu=(u_1,..., u_p)'$, $\|\bu\|_\infty=\max_i |u_i|$, and $\bI_k$ denotes the $k\times k$ identity matrix. For a matrix $\bH=(h_{ij})$,  $\|\bH
\|_2=\sqrt{\lambda_{\max} (\bH' \bH ) }$ is the operator norm, where
$\lambda_{\max} (\cdot) $ denotes for the largest eigenvalue of a matrix, and $\|\bH\|_{\min}$ is the square root of the minimum non-zero eigenvalue of $\bH'\bH$. The superscript $'$ denotes 
the transpose of a vector or matrix. 
Finally, we use the notation $a\asymp b$ to denote $a=O(b)$ and $b=O(a)$.

\section{The Model and Methodology}
In this section, we state the proposed model, discuss the estimation methodology, and 
an estimation procedure when the number of common factors is known. This is followed 
by the proposed method for determining the number of common factors. 
We also provide a flowchat to summarize the proposed modeling procedure. 

\subsection{Setting}
Let $\by_t=(y_{1t},...,y_{pt})'$ be an observable $p$-dimensional time series. We assume $E(\by_t) = {\bf 0}$ and $\by_t$ admits a latent structure:
\begin{equation}\label{decom}
\by_t=\bL\left[\begin{array}{c}
\bff_t\\
\bve_t\\
\end{array}\right]=[ \bL_1, \bL_2] \left[\begin{array}{c} \bff_t \\ \bve_t\end{array}\right] = \bL_1\bff_t+\bL_2\bve_t,
\end{equation}
where $\bL\in R^{p\times p}$ is a full rank loading matrix, $\bff_t = (f_{1t},\ldots,
f_{rt})'$ is a $r$-dimensional dynamically dependent factor process, 
$\bve_t = (\varepsilon_{1t},\ldots,\varepsilon_{vt})'$ is a $v$-dimensional white noise vector, 
and $r+v = p$.  For meaningful dimension reduction, we assume $r$ is a small fixed 
nonnegative integer. 
In addition, we also assume $\Cov(\bff_t)=\bI_r$, $\Cov(\bve_t)=\bI_v$, 
$\Cov(\bff_t,\bve_t)={\bf 0}$, and no linear combination of $\bff_t$ is serially uncorrelated.  The last assumption is trivial, because we can reduce the dimension of 
$\bff_t$ if any such a linear combination exists. The assumptions on the covariance 
matrices of $\bff_t$ and $\bve_t$ are common among factor models. 
The decomposition of Model (\ref{decom}) is general in the sense that any  
finite-order VARMA time series $\by_t$ can always be written in Equation (\ref{decom}) 
via canonical correlation analysis between  constructed vectors involving $\by_t$ and its 
lagged values. See \cite{TiaoTsay_1989}.  Readers are referred to 
\cite{anderson2003} for details of canonical correlation analysis. 

If $v=0$, then no linear combination of $\by_t$ is white noise, indicating no dimension 
reduction. Therefore, we focus on the case of $v > 0$ in this paper. 
Under Equation (\ref{decom}), the dynamic dependence of $\by_t$  
is driven by $\bff_t$ if $r > 0$. Thus, $\bff_t$ indeed consists of the common dynamically dependent factors of $\by_t$. In the terminology of \cite{TiaoTsay_1989}, 
(a) $\bve_t$ is a $v$-dimensional scalar component process of 
order (0,0) if $v > 0$, that is, there exists a transformation matrix $\bV_2\in R^{p\times v}$ such that $\bve_t=\bV_2'\by_t$ and \Cov($\bve_t,\by_{t-j}$) = $\bzero$ for $j > 0$, 
and (b) no linear combination of $\bff_t$ is a scalar component of order (0,0) if $r > 0$. 
Readers are referred to \cite{TiaoTsay_1989} 
for a formal definition of a scalar component of order (0,0).  
Condition (a) is equivalent to $\bve_t$ 
being a white noise under the traditional factor models, where $\bff_t$ and 
$\bve_t$ are assumed to be independent. 

Assuming that the time series $\by_t$ follows 
 a structural model consisting of trend, seasonal component, 
and irregular series,  \cite{gaotsay2018} employ Equation (\ref{decom})
to model the irregular series. These authors use CCA to determine 
the number of common factors. 
 However, the method of CCA only works when $p = o(n^{1/2})$, where $n$ is the 
 sample size. This restricts the applicability of the modeling procedure of 
 \cite{gaotsay2018}. We relax such restrictions on $p$ in this paper. 

To study Model (\ref{decom}) in a more general setting and to provide sufficient statistical insights on the proposed factor models, we decompose $\bL_1$ and $\bL_2$ of Model (\ref{decom}) as $\bL_1=\bA_1\bQ_1$ and $\bL_2=\bA_2\bQ_2$, respectively, where $\bA_1$ and $\bA_2$ are two half orthonormal  matrices, i.e., $\bA_1'\bA_1=\bI_r$ and $\bA_2'\bA_2=\bI_v$. This can be done via the QR decomposition or singular value decomposition, and hence most of the strengths of $\bL_1$ and $\bL_2$ are retained in $\bQ_1$ and $\bQ_2$, respectively. Furthermore, let $\bx_t=\bQ_1\bff_t$ and $\be_t=\bQ_2\bve_t$, then Model (\ref{decom}) can be written as
\begin{equation}\label{sfactor}
\by_t=\bA_1\bx_t+\bA_2\be_t, 
\end{equation}
which is close to the traditional factor model in Equation (\ref{tra-f}). 
Some remarks are in order. First, the model we are investigating is still the one in (\ref{decom}) and we rewrite it in the form of (\ref{sfactor}) to provide more insights to the proposed methods below. Second, even though $\bL$ is of full rank, $\bA_1$ is not orthogonal to $\bA_2$ in general because we performed the decomposition separately. Third, $\bA_1$ and $\bx_t$ are still not uniquely identified because 
we can replace $(\bA_1,\bx_t)$ by $(\bA_1\bH,\bH'\bx_t)$ for any orthonormal matrix 
$\bH\in R^{r\times r}$, where $\bA_1$ and $\bA_1\bH$ are both half orthonormal matrices, and it is also due to that the decomposition of $\bL_1$ is not unique. The same issue applies to $\bA_2$ and $\be_t$. Nevertheless, the linear space spanned by the columns of $\bA_1$, denoted
by $\mathcal{M}(\bA_1)$, is uniquely defined, and is equal to that of $\bL_1$. $\mathcal{M}(\bA_1)$ is called the factor loading space,  and the linear space $\mathcal{M}(\bA_2)$  can be defined similarly for the idiosyncratic component.

\subsection{Estimation Methodology}

To begin, we provide some rationale for the proposed estimation method.  
Let $\bB_1$ and $\bB_2$ be the orthonormal complement of $\bA_1$ and $\bA_2$, respectively, 
i.e., $\bB_1\in R^{p\times v}$ and $\bB_2\in R^{p\times r}$ are two half orthonormal matrices 
satisfying $\bB_1'\bA_1={\bf 0}$ and $\bB_2'\bA_2={\bf 0}$. Denote 
$[\bA_1,\bB_1]=[\ba_1,...,\ba_r,\bb_1,...,\bb_v]$ and $[\bA_2,\bB_2]=[\ba_{r+1},...,\ba_p,\bb_{v+1},...,\bb_p]$, which are $p\times p$ matrices. It follows from Model (\ref{sfactor}) that 
\begin{equation}\label{t-white}
\bB_1'\by_t=\bB_1'\bA_2\be_t,
\end{equation}
and, hence, $\bB_1'\by_t$ is a $v$-dimensional white noise process. In other words, for any column 
$\bb_j$ of $\bB_1$ with $1\leq j\leq v$, $\{\bb_j'\by_t, t=0,\pm1,\ldots \}$ is a white noise process. 

Unlike the traditional factor models, which assume $\bx_t$ and $\be_s$ are uncorrelated for 
any $t$ and $s$,  we only require $\Cov(\bx_t,\be_{t+j})={\bf 0}$ for $j\geq { 0}$ in this paper.  
For $k \geq 0$, let 
\[\bSigma_y(k)=\Cov(\by_t,\by_{t-k}),\,\,\bSigma_x(k)=\Cov(\bx_t,\bx_{t-k}),\,\,\bSigma_{xe}(k)=\Cov(\bx_t,\be_{t-k}),\]
be the covariance matrices of interest. 
It follows from (\ref{sfactor}) that
\begin{equation}\label{sigy-r}
\bSigma_y(k)=\bA_1\bSigma_x(k)\bA_1'+\bA_1\bSigma_{xe}(k)\bA_2', \quad k\geq 1,
\end{equation}
and, for $k=0$, 
\begin{equation}\label{covy}
\bSigma_y \equiv \bSigma_y(0) =\bA_1\bSigma_x\bA_1'+\bA_2\bSigma_e\bA_2'.
\end{equation}
For a pre-specified integer $k_0>0$, define
\begin{equation}\label{M}
\bM=\sum_{k=1}^{k_0}\bSigma_y(k)\bSigma_y(k)',
\end{equation}
which is a $p \times p$ semi-positive definite matrix. 
By $\bB_1'\bA_1={\bf 0}$, we have 
$\bM\bB_1={\bf 0}$, that is, the columns of $\bB_1$ are the eigenvectors associated with the zero eigenvalues of $\bM$, and the factor loading space $\mathcal{M}(\bA_1)$ is spanned by the eigenvectors associated with the $r$ non-zero eigenvalues of $\bM$. Note that the form of $\bM$ in Equation (\ref{M}) is a special case of the Orthonormalized Partial Least Squares of time series data. See the discussion in Section 5.
For $k_0 > 1$, the summation in the definition of $\bM$ enables us to pool information 
over different lags, which is particularly helpful when the sample size is small. In practice, with 
a given sample size, the estimation accuracy of auto-covariance matrices of $\by_t$ deteriorates 
as the lag $k$ increases. Thus, some compromise in selecting $k_0$ is needed in real applications. 
Limited experience suggests that a relatively small $k_0$ is sufficient in providing useful information 
concerning the  model structure of $\by_t$, because, for a stationary time series, cross-correlation 
matrices decay to zero exponentially as $k$ increases. 
Also, the choice of $k_0$ seems to be not sensitive. 
See, for instance, the simulation results in Section 4 and the online supplement.

Turn to the estimation of the common factors. We observe that, from Equation (\ref{sfactor}), 
\begin{equation}\label{r-factor}
\bB_2'\by_{t}=\bB_2'\bA_1\bx_{t},
\end{equation}
which  is uncorrelated with $\bB_1'\by_t$ defined in (\ref{t-white}). Therefore, 
\begin{equation}\label{b2}
\bB_2'\bSigma_y\bB_1\bB_1'\bSigma_y\bB_2={\bf 0},
\end{equation}
which implies that $\bB_2$ consists of the last $r$ eigenvectors corresponding to the zero eigenvalues of $\bS:=\bSigma_y\bB_1\bB_1'\bSigma_y$. From the relationship in (\ref{r-factor}) and the discussion of Remark 1 in Section 3 below, $\bB_2'\bA_1$ is a $r\times r$ invertible matrix and hence $\bx_t=(\bB_2'\bA_1)^{-1}\bB_2'\by_t$. 
From Equation (\ref{sfactor}),  $\bx_t$ does not include the white noise term. 
Moreover, the columns in $\bA_2$ can be treated as the eigenvectors associated with the non-zero eigenvalues of $\bS$. Finally, even though   $\bB_2$ (also $\bA_2$) is not unique and 
$\bB_2\bH$ is also a solution to (\ref{b2}) for any orthonormal matrix $\bH\in R^{r\times r}$, 
this non-uniqueness does not alter the representation of $\bx_t=(\bB_2'\bA_1)^{-1}\bB_2'\by_t$. 

\subsection{Estimation When the Number of Common Factors is Known}

For a given realization $\{\by_t|t=1, \ldots,n\}$, we discuss our estimation 
procedure in this section, assuming that the number of common factors $r$ is known. 
The selection of $r$ is discussed in the next section. 
Let $\wh\bSigma_y(k)$ be the lag-$k$ sample auto-covariance matrix of $\by_t$. To estimate $\mathcal{M}(\bA_1)$, we perform an eigen-analysis of 
\begin{equation}\label{mhat}
\wh\bM=\sum_{k=1}^{k_0}\wh\bSigma_y(k)\wh\bSigma_y(k)',
\end{equation}
where, as before, $k_0$ is a pre-specified positive integer. 
Let $\wh\bA_1=[\wh\ba_1,...,\wh\ba_r]$ and $\wh\bB_1=[\wh\bb_1,...,\wh\bb_v]$ be two half orthonormal matrices consisting of the eigenvectors of  $\wh\bM$ corresponding to the non-zero 
and zero eigenvalues, respectively. In view of Equation (\ref{b2}), we next perform 
another eigen-analysis of  
\begin{equation}\label{cs}
\wh\bS=\wh\bSigma_y\wh\bB_1\wh\bB_1'\wh\bSigma_y,
\end{equation}
which is a projected PCA. That is, we project the data $\by_t$ onto the direction of $\wh\bB_1$, then perform the PCA between the original data $\by_t$ and its projected coordinates. 
Note that $\wh\bS\in R^{p\times p}$ and its rank is at most $p-r$.
If the dimension $p$ is small, we employ $\wh\bB_2=[\wh\bb_{v+1},...,\wh\bb_p]$, where  $\wh\bb_{v+1},...,\wh\bb_p$ are the eigenvectors corresponding to the smallest $r$ 
eigenvalues of $\wh\bS$. 
On the other hand, if 
$p$ is relatively large, we expect the covariance of the idiosyncratic terms also captures some of the cross-sectional dependence of the data as the covariance of the factors does. That is, given a high-dimensional covariance matrix $\bSigma_e$, we entertain 
that the largest $K$ eigenvalues of $\bSigma_e$ are diverging. Therefore, 
we write $\bA_2=[\bA_{21},\bA_{22}]$ with $\bA_{21}\in R^{p\times K}$ 
and $\bA_{22}\in R^{p\times (v-K)}$ and consider the linear space $\mathcal{M}(\bB_2^*)$, 
where $\bB_2^*=[\bA_{22},\bB_2] \in R^{p\times (p-K)}$.
Note that $\bB_2^*$ consists of $p-K$ eigenvectors 
corresponding to the  $p-K$ smallest eigenvalues of $\bS=\bSigma_y\bB_1\bB_1'\bSigma_y$ 
defined before. 
Let $\wh{\bB}_2^*$ be an estimator of $\bB_2^*$ consisting of the $p-K$ eigenvectors 
associated with the $p-K$ smallest eigenvalues of $\wh\bS$. We then estimate  $\wh\bB_2$ by  $\wh\bB_2=\wh{\bB}_2^*\wh\bR$, where $\wh\bR=[\wh\br_1,\ldots,\wh\br_r]\in R^{(p-K)\times r}$ with $\wh\br_i$ being the eigenvector associated with the $i$-th largest eigenvalues of ${\wh{\bB}_2^{*}{'}}\wh{\bA}_1\wh{\bA}_1'{\wh{\bB}_2^*}$. 
This choice of estimator guarantees that the matrix $(\wh\bB_2'\wh{\bA}_1)^{-1}$ behaves well 
in recovering the common factor $\wh\bx_t$. Although, in this case, $\wh\bB_2$ is not a consistent estimator for $\bB_2$, but $\wh\bB_2^*$ is a consistent one for $\bB_2^*$, which is sufficient to eliminate the diverging effect of $\bSigma_{e}$. 
Detailed properties of the estimators are given in Section 3.
Finally, we recover the factor process as $\wh\bx_t=(\wh\bB_2'\wh\bA_1)^{-1}\wh\bB_2'\by_t$.

\subsection{Determination of the Number of Common Factors}
The estimation of $\bA_1$ and $\bx_t$ in the prior sections is based on a given $r$, which 
is unknown in practice. 
There are several methods available in the literature to determine $r$ for the traditional factor model in Equation (\ref{tra-f}). See, for example, the information criterion in \cite{BaiNg_Econometrica_2002} and \cite{Bai_Econometrica_2003}, the random matrix theory method in \cite{onatski2010}, and the ratio-based method in Lam and Yao (2012),  \cite{ahn2013}, and \cite{lietal2017}, among others. However, none of these methods is applicable to  Model (\ref{decom}) directly. 
The most relevant method is the one based on testing the number of zero canonical correlations between $\by_t$ and an extended vector of its lagged values employed in \cite{gaotsay2018}. But this 
testing method  only works when the dimension $p$ is relatively small with respect to 
the sample size $n$. For a large $p$, some alternatives must be sought. 

In this section, we propose a new approach to estimate the number of common factors 
based on Equation (\ref{t-white}). Specifically, we perform white noise tests to determine the number of white noise components $\hat{v}$ and use $\hat{r}=p-\hat{v}$. 
Let  $\wh\bG$ be the matrix of eigenvectors (in the decreasing order of 
eigenvalues) of the sample matrix $\wh\bM$ of Equation (\ref{mhat}) 
and $\wh\bu_t=\wh\bG'\by_t$ = $(\hat{u}_{1t}, \ldots, \hat{u}_{pt})'$ 
be the transformed series. We propose to test sequentially the 
number of white noises in $\wh\bu_t$, which is an estimate of $v$. To this end, 
we consider two cases depending on the dimension $p$.

If the dimension $p$ is small, we recommend using a bottom-up procedure to determine 
the number of white noise components. Specifically, we use the conventional test statistics, 
such as the well-known Ljung-Box statistic $Q(m)$ or its rank-based variant, to test the null hypothesis that $\hat{u}_{it}$ is  
a white noise series starting with $i = p$. 
If the null hypothesis is rejected, then $\hat{v} = 0$ and $\hat{r} = p$; 
otherwise, reduce $i$ by one and repeat the testing process. 
Clearly, this testing process can only last until $i=1$. 
If all transformed series $\hat{u}_{it}$ are white noise, then $\hat{v} = p$ and 
$\hat{r} = 0$. In general, if $\hat{u}_{it}$ is not a white noise series but $\hat{u}_{jt}$ are for $j = i+1, \ldots, p$, then  $\hat{v} = p-i$ and $\hat{r} = i$, and we have 
$\wh\bG$ = $[\wh\bA_1,\wh\bB_1]$, where $\wh\bA_1\in R^{p\times \wh r}$ and $\wh\bB_1\in R^{p\times \wh v}$.

For a large $p$, the conventional white-noise test statistics are no longer adequate, but 
some methods  have been developed in recent years to test high-dimensional white noise. 
See, for instance,  \cite{changyaozhou2017}, \cite{Tsay_2018}, and \cite{lietal2019}. 
We only consider the first two methods in this paper since the third one requires the noise to have an independent structure and its covariance is identity. The method by \cite{changyaozhou2017} makes use of the maximum absolute auto-correlations and cross-correlations of the component series. Specifically, let $\bw_t$ = $(w_{1t}, \ldots, w_{dt})'$ 
be a $d$-dimensional real-valued time series. In this paper, $1 \leq d \leq p$. 
Define the lag-$k$ sample 
covariance matrix as  
 $\wh\bSigma_{w}(k)=(n-k)^{-1}\sum_{t=k+1}^n(\bw_{t}-\bar{\bw})(\bw_{t-k}-\bar{\bw})',$ where $\bar{\bw}=n^{-1}\sum_{t=1}^n\bw_{t}$ is the sample mean.  
 The test statistic $T_n$ of \cite{changyaozhou2017} is 
\begin{equation}\label{t:test}
T_n=\max_{1\leq k\leq \bar{k}} T_{n,k},
\end{equation}
where $\bar{k}\geq 1$ is a pre-specified positive integer and $T_{n,k}=\max_{1\leq j,l\leq d}n^{1/2}|\wh\rho_{jl}(k)|$ with
\[\wh\bGamma_{w}(k)\equiv\left[\wh\rho_{jl}(k)\right]_{1\leq j,l\leq d}=\diag\{\wh\bSigma_{w}(0)\}^{-1/2}\wh\bSigma_{w}(k)\diag\{\wh\bSigma_{w}(0)\}^{-1/2}.\]
The limiting distribution of $T_n$ can be approximated by that of the $L_{\infty}$-norm of a normal random vector, i.e., there exists a  random variable $\bz_d \sim N({\bf 0}, \bTheta_{d,n})$ such that
\[\sup_{s\geq 0}|P(T_n>s)-P(\|\bz_d\|_{\infty}>s)|=o(1),\]
where $\bTheta_{d,n}$ is close to the asymptotic covariance of the vector containing the columns of $\wh\bGamma_{w}(1)$ to $\wh\bGamma_{w}(\bar{k})$, and it can be estimated from $\{\bw_{t}|t=1,...,n\}$. Therefore, the critical values of $T_n$ 
can be obtained by bootstrapping from a multivariate normal distribution.

The second method of high-dimensional white noise test is introduced by \cite{Tsay_2018} 
using the extreme value theory. The test is robust with a closed-form limiting distribution under 
some weak assumptions and is easy to use in practice. The basic idea of the test is as follows.
Consider a $d$-dimensional time series $\bw_t$ with a realization of $n$ observations 
$\{\bw_t| t=1, \ldots, n\}$. Assume, for now, that $d < n$. 
Let $\wt\bw_{t}$ =  $\bSigma_{w}^{-1/2}\bw_{t}$ be a standardized series, where $\bSigma_{w}^{1/2}$ is a square-root matrix of  the covariance matrix $\bSigma_{w}$. 
With $d < n$, this standardization can be done by PCA.  
For simplicity, we denote the standardized realization as $\wt\bw_{t}$ =  $\wh \bSigma_{w}^{-1/2}\bw_{t}$. If $d \geq n$, $\wh\bSigma_w$ is singular and we 
discuss a modification later. 
Note that the components of $\wt\bw_{t}=(\wt w_{1t}, \ldots, \wt w_{dt})'$  are mutually uncorrelated.  Next, let $\wh{\boldsymbol{\varrho}}_{t}=(\wh \varrho_{1t}, \ldots, \wh \varrho_{dt})'$ be the rank series 
of $\wt\bw_t$, where $\wh \varrho_{jt}$ is the rank of $\wt w_{jt}$ in $\{\wt w_{j,1},...,\wt w_{j,n}\}$ for $1 \leq j \leq d$. The lag-$\ell$ rank cross-correlation matrix is then defined as
\[\wh\bGamma_{w,\ell}=\frac{12}{n(n^2-1)}\sum_{t=\ell+1}^n(\wh{\boldsymbol{\varrho}}_{t}-\bar{{\boldsymbol{\varrho}}})(\wh{\boldsymbol{\varrho}}_{t-\ell}-\bar{{\boldsymbol{\varrho}}})',\]
where $\bar{{\boldsymbol{\varrho}}}=\frac{n+1}{2}{\bf 1}_d$ and ${\bf 1}_d$ is a $d$-dimensional vector of ones.  The test statistic of \cite{Tsay_2018} for testing that there is no serial or cross-sectional 
correlation in the first $m$ lags of $\bw_t$ is 
\begin{equation}\label{Tsay:test}
T(m)=\max\{\sqrt{n}|\wh\bGamma_{w,\ell}(j,k)|:1\leq j,k\leq d,\ 1\leq \ell\leq m\},
\end{equation}
where $\wh\bGamma_{w,\ell}(j,k)$ is the $(j,k)$-th element of $\wh\bGamma_{w,\ell}$. 
By the extreme-value theory, the limiting distribution of $T(m)$ under the white noise hypothesis 
is a Gumbel distribution provided that the component series of $\bw_t$ follow  a continuous distribution.  Therefore, we reject the null hypothesis
$H_0: \bw_{t} \,\,\text{is a vector white noise}$
at the $\alpha$-level if 
\[T(m)\geq c_{d,m}\times x_{1-\alpha/2}+s_{d,m},\]
where $x_{1-\alpha/2}=-\log(-\log(1-\alpha/2))$ is the $(1-\alpha/2)$-th quantile of the 
standard Gumbel distribution and
\[c_{d,m}=[2\log(d^2m)]^{-1/2},\,\,\text{and}\,\, s_{d,m}=\sqrt{2\log(d^2m)}-\frac{\log(4\pi)+\log(\log(d^2m))}{2(2\log(d^2m))^{1/2}}.\]

If $d \geq n$,  the sample covariance matrix of $\bw_{t}$ is singular and some alternative 
methods must be sought to create mutually uncorrelated series. \cite{Tsay_2018} 
provided a 
method  by selecting a subset series of $\bw_t$ to perform testing, and the method 
works reasonably well in simulations and some illustrative applications. 
In this paper, we consider a simpler method 
by using the relations in (\ref{sfactor}) and (\ref{t-white}). Note that in our testing, $\bw_t$ is a subset 
of the transformed series $\wh \bu_t = \wh\bG'\by_t$. 
Since $\wh\bM$ is based on the covariance 
matrices of $\by_t$ and its lagged values the components of $\wh \bu_t$ associated with 
small eigenvalues contain little information on the 
dynamical dependence of $\by_t$. Therefore, we can drop the last $(p-\epsilon n)$ components 
of the transformed series $\wh \bu_t$ without affecting the white-noise test, where $\epsilon \in (0,1)$. In other words, when $p > n$, we cannot start with $\bw_t = \wh\bu_t$, 
but we can choose  $\bw_t$ to consist of the first $d = \epsilon n < n$ components of 
$\wh \bu_t$ to perform the white-noise test without affecting the determination of $r$ under 
the assumption that $r$ is small in applications. 

Return to the determination of $r$ when $p$ is large. We can apply  the 
high-dimensional white noise test of  \cite{changyaozhou2017} or \cite{Tsay_2018} 
to subsets of the transformed series $\wh \bu_t$. We choose the subset simply by dropping the last $p-\epsilon n$ projected coordinates in $\wh\bu_t$ when $p\geq n$ because those coordinates capture less dynamic dependencies by the nature of the projection method. 
Specifically, let $p_* = p$ if $p < n$ and $p_* = \epsilon n $  if $p \geq n$, 
where $\epsilon \in (0,1)$. 
Starting with $i = 1$ and $\bw_t = (\hat{u}_{i,t}, \ldots, \hat{u}_{p_*,t})'$, 
we test the null hypothesis that $\bw_t$ has no serial or cross-sectional correlations in the 
first $m$ lags using a selected test statistic. If the null hypothesis is rejected, increase 
$i$ by one and repeat the testing process. Using this testing process, we select 
$\hat{r}$ as $i-1$ for which the $i$th test does not reject the null hypothesis. 
Note that since both test statistics considered use the maximum of absolute correlations, 
the computation of the testing process is trivial because we only need to compute 
the cross-correlation matrices of $\bw_t$ at each time.

\subsection{Modeling Procedure}

Figure~\ref{fig0} provides a flowchat that summarizes the proposed modeling procedure.
Selecting a small positive integer $k_0$, we start with an eigenvalue-eigenvector analysis 
of the sample matrix $\wh\bM$, then apply the white-noise test to the 
transformed series, using the eigenvectors of $\wh\bM$, to determine the dimension 
$\hat{v}$ of 
the white noise process, which in turn provides an estimate of the number of 
common factors $\hat{r} = p-\hat{v}$. Next, we perform the proposed projected 
PCA to find the orthogonal directions of the diverging noises so that the common 
factors can be properly estimated. The projected PCA can mitigate the impact of 
white noises in recovering the common factors. 
Finally, 
with $\wh\bA_1$ and the estimated factor process $\wh\bx_t$, we compute the $h$-step ahead prediction of the $\by_t$ series using the formula $\wh\by_{n+h}=\wh\bA_1\wh\bx_{n+h}$, where $\wh\bx_{t+h}$ is an $h$-step ahead forecast for $\bx_t$ based on the estimated past values $\wh\bx_1,\ldots, \wh\bx_n$. This can be done, for example, by fitting a  VAR model 
to $\{\wh\bx_1, \ldots, \wh\bx_n\}$. Alternatively, we may also adopt the diffusion index models to do forecasting if we are interested in some particular component of $\by_t$. See \cite{StockWatson_2002a} and \cite{StockWatson_2002b} for details.


\tikzstyle{decision} = [diamond, draw, fill=blue!20, 
    text width=5em, text badly centered, node distance=3cm, inner sep=0pt]
\tikzstyle{block} = [rectangle, draw, fill=blue!20, 
    text width=11.5em, text centered, rounded corners, minimum height=4em]
\tikzstyle{mycircle} = [circle, thick, draw=orange, minimum height=4mm]

\tikzstyle{line} = [draw, -latex']
\tikzstyle{cloud} = [draw, ellipse,fill=red!20, node distance=3cm,
    minimum height=2em]
  \begin{figure}[h]
\begin{center}

     \begin{tikzpicture}[align=center,node distance = 2cm, auto]
    \node [block] (init) {Input the data:\\
$\{{\bf y}_t\in R^{p}: 1\leq t\leq n\}$};
    \node [block, right of=init, node distance=10cm] (trip) {Obtain the eigenvector matrix $\hat{\bf G}=[\hat{\bf A}_1,\hat{\bf B}_1]\in R^{p\times p}$};
\node [block, below of=trip, node distance=4cm] (wtest) {Identify $\hat{r}$ such that $\hat{\bf A}_1\in R^{p\times \hat{r}}$, $\hat{\bf B}_1\in R^{p\times (p-\hat{r})}$, and $\hat{\bf B}'y_t$ is a vector white noise};
\node [block, below of=init, node distance=4cm](idk) {Identify $\hat{K}$ via (3.4), and $\hat{\bf B}_2^*\in R^{p\times (p-\hat{K})}$ consists of the eigenvectors of $\hat{\bf S}$ associated with the $(p-\hat{K})$ smallest eigenvalues};
\node[block, below of=idk, node distance=6cm] (b2hat) {Obtain $\hat{\bf B}_2=\hat{\bf B}_2^*{\bf R}$, where ${\bf R}$ consistes of the eigenvectors associated with the $\hat{r}$ largest eigenvalues of $\hat{\bf B}_2^*{'}\hat{\bf A}_1\hat{\bf A}_1'\hat{\bf B}_2^*$};
\node [block,below of=wtest, node distance=6cm](output){Output the estimated factors $\hat{\bf x}_t=(\hat{\bf B}_2'\hat{\bf A}_1)^{-1}\hat{\bf B}_2'{\bf y}_t$, the factor loading matrix $\hat{\bf A}_1$, and the white noise components ${\bf y}_t-\hat{\bf A}_1(\hat{\bf B}_2'\hat{\bf A}_1)^{-1}\hat{\bf B}_2'{\bf y}_t$};

   \path [line] (init) --node {Eigenanalysis on $\hat{\bf M}$ of (\ref{mhat})}(trip);

 \path [line] (trip) --node {\textcircled{1}}(wtest);
 \path [line] (wtest) --node  {Projected PCA on $\hat{\bf S}$ in (\ref{cs})}(idk);
  \path [line] (idk) --node {\textcircled{2}}(b2hat);
  \path [line] (b2hat) --node {output}(output);
\end{tikzpicture}
\caption{The flowchart of the proposed modeling procedure. \textcircled{1}:  Applying sequential vector white noise test on $\hat{\bf G}'{\bf y}_t$; \textcircled{2}: Finding the orthogonal directions of the diverging noises.}\label{fig0}
  \end{center}
  \end{figure}
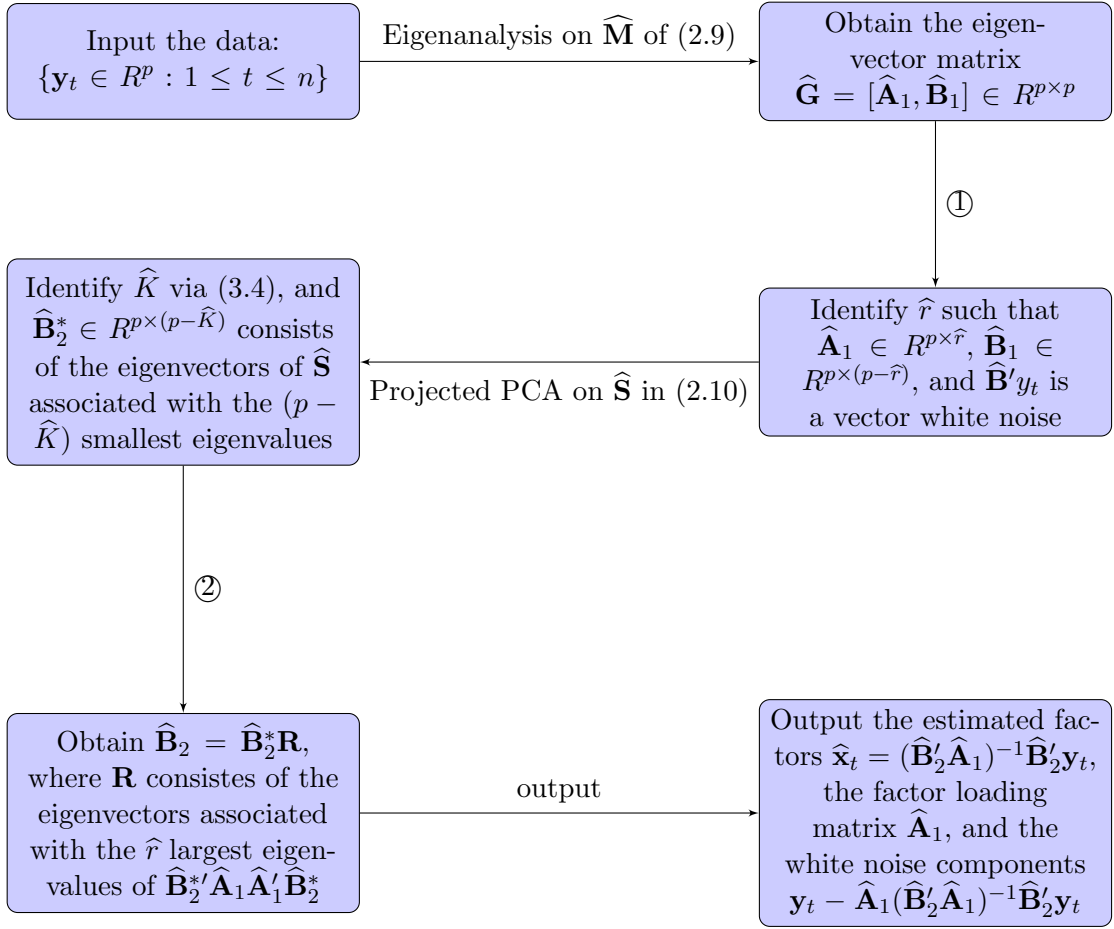
  

\section{Theoretical Properties}
This section studies the asymptotic theory of the estimation method used in the paper. 
We start with the assumption that the number of common factors $r$ is known, and  
divide the derivations into two cases depending on the value of the dimension $p$. 
The case of estimated $r$ is discussed later.

\subsection{Asymptotic Properties When $p$ is Fixed, But $n\rightarrow\infty$}

We consider first the asymptotic properties of the estimators when $p$ is fixed but 
$n\rightarrow\infty$. These properties show the behavior of our estimation method 
when $n$ is large and $p$ is relatively small. We begin with the assumptions used. 

\begin{assumption}
The process $\{(\by_t,\bff_t)\}$ is $\alpha$-mixing with the mixing coefficient satisfying the condition $\sum_{k=1}^\infty\alpha_p(k)^{1-2/\gamma}<\infty$ for some $\gamma>2$, where
\[\alpha_p(k)=\sup_{i}\sup_{A\in\mathcal{F}_{-\infty}^i,B\in \mathcal{F}_{i+k}^\infty}|P(A\cap B)-P(A)P(B)|,
\]
and $\mathcal{F}_i^j$ is the $\sigma$-field generated by $\{(\by_t,\bff_t):i\leq t\leq j\}$.
\end{assumption}
\begin{assumption}
$E|f_{it}|^{2\gamma}<C_1$ and $E|\ve_{jt}|^{2\gamma}<C_2$ for $1\leq i\leq r$ and $1\leq j\leq v$, where $C_1$ and $C_2>0$ are constants, and $\gamma$ is given in Assumption 1.
\end{assumption}
\begin{assumption}
$\lambda_1>...>\lambda_r>\lambda_{r+1}=...=\lambda_p=0$, where $\lambda_i$ is the $i$-th largest eigenvalue of the $\bM$ matrix in Equation (\ref{M}).
\end{assumption}
Assumption 1 is standard for dependent random processes. See \cite{gaoetal2017} for a theoretical justification for VAR models. The conditions in Assumption 2 imply that $E|y_{it}|^{2\gamma}< \infty$ under the setting that $p$ is fixed. The moment condition used is slightly stronger than the ones in \cite{lamyao2012} and \cite{fan2013}, which only assume the existence of finite fourth moments,  because we use the $\alpha$-mixing condition on the series, which is weaker than the $\psi$-mixing in \cite{lamyao2012} and the exponential  rate of decay of the mixing coefficient in \cite{fan2013}. In Assumption 3, if the $r$ non-zero eigenvalues of $\bM$ are distinct, the eigenvector matrix 
$\bA_1$ is uniquely defined if we ignore the trivial replacement of $\ba_j$ by $-\ba_j$ for $1\leq j\leq r$. In general, the choice of $\bA_1$ in Model (\ref{sfactor}) is not
unique without Assumption 3, so we consider the error in estimating $\mathcal{M}(\bA_1)$, the column space of $\bA_1$, because $\mathcal{M}(\bA_1)$ is uniquely defined by
(\ref{sfactor}) and it does not vary with different choices of
$\bA_1$. The same argument also applies to matrices $\bA_2$, $\bB_1$, and $\bB_2$. 
To this end, we adopt the discrepancy measure used by
\cite{panyao2008}: for two $p\times r$ half orthogonal
matrices ${\bf H}_1$ and ${\bf H}_2$ satisfying the condition ${\bf
H}_1'{\bf H}_1={\bf H}_2'{\bf H}_2=\bI_{r}$, the difference
between the two linear spaces $\mathcal{M}({\bf H}_1)$ and
$\mathcal{M}({\bf H}_2)$ is measured by
\begin{equation}
D({\bf H}_1,{\bf
H}_2)=\sqrt{1-\frac{1}{r}\textrm{tr}({\bf H}_1{\bf H}_1'{\bf
H}_2{\bf H}_2')}.\label{eq:D}
\end{equation}
Note that $D({\bf H}_1,{\bf H}_2) \in [0,1].$ 
It is equal to $0$ if and only if
$\mathcal{M}({\bf H}_1)=\mathcal{M}({\bf H}_2)$, and to $1$ if and
only if $\mathcal{M}({\bf H}_1)\perp \mathcal{M}({\bf H}_2)$.  
The following theorem establishes the consistency of the estimated loading matrix $\wh\bA_1$, its orthonormal complement $\wh\bB_1$, the matrix $\wh\bB_2$, and the extracted common factor $\wh\bA_1\wh\bx_t$.
\begin{theorem}\label{tm1}
Suppose Assumptions 1 and 2 hold and $r$ is known and fixed. Then, for fixed $p$, 
\[D(\wh\bA_1,\bA_1)=O_p(n^{-1/2}),\quad D(\wh\bB_1,\bB_1)=O_p(n^{-1/2}),\quad\text{and}\quad D(\wh\bB_2,\bB_2)=O_p(n^{-1/2}),\]
as $n\rightarrow\infty$. If Assumption 3 also holds, we further have $\|\wh\bA_1-\bA_1\|_2=O_p(n^{-1/2})$.
Therefore,
\[\|\wh\bA_1\wh\bx_t-\bA_1\bx_t\|_2=O_p(n^{-1/2}).\]
\end{theorem}
\begin{remark}
From Theorem~\ref{tm1} and, as expected, the convergence rates of all estimates 
are standard at $\sqrt{n}$, which is commonly seen in the traditional statistical theory. 
To recover the factor process, we need to guarantee that $\bB_2'\bA_1$ is invertible. This follows from the fact that there exist $\bR_1\in R^{r\times r}$ and $\bR_2\in R^{v\times r}$ such that $\bB_2=\bL_1\bR_1+\bL_2\bR_2=\bA_1\bQ_1\bR_1+\bA_2\bQ_2\bR_2$, i.e., each column of $\bB_2$ can be represented as a linear combination of the columns of $\bL$. Therefore, $\bI_r=\bB_2'\bB_2=\bB_2'\bA_1\bQ_1\bR_1$ and, hence, $rank(\bB_2'\bA_1)=r$, which is of full rank.
\end{remark}

Theorem~\ref{tm1} implies that the convergence rate does not change even when some 
non-zero eigenvalues of $\bM$ are not distinct and $\bA_1$ is not uniquely defined. 
In fact, 
the consistency of the column spaces $\mathcal{M}(\bB_1)$ and $\mathcal{M}(\bB_2)$ is more meaningful because their columns correspond to the zero eigenvalues of $\bM$ and $\bS$, respectively, and cannot be uniquely characterized.

\subsection{Asymptotic Properties When $n\rightarrow\infty$ And $p\rightarrow\infty$}

Turn to the case of high-dimensional time series. Note that our framework is different from the PCA approach in \cite{BaiNg_Econometrica_2002} and \cite{fan2013}, where the consistency of the leading eigenvectors  does not require any restriction between $p$ and $n$ by using the well-known $\sin(\theta)$ theorem of \cite{davis1970} and \cite{yuetal2014}, among others, because they imposed the pervasiveness assumption that all eigenvalues associated with the factors and loadings are diverging at the rate of $p$, which implies that the covariance has some spiked eigenvalues. 
See Section 2 of \cite{fan2013} for details. However, certain developments in random matrix theory have shown that the true eigenvalues and eigenvectors might not be consistently estimated from the sample covariance matrix when $p/n$ is not negligible in general. See, for example, \cite{johnstone-lu2009} and \cite{paul2007}. Therefore, using sample auto-covariance-based analysis without the pervasiveness assumption, we do expect some theoretical restrictions between $p$ and $n$. 
It is well-known that if the dimension $p$ 
diverges faster than $n^{1/2}$, the sample covariance matrix is no longer a consistent 
estimate of the population covariance matrix. 
When $p=o(n^{1/2})$, it is still possible to consistently estimate the factor loading matrix $\bA$ and the number of common factors $r$. See \cite{gaotsay2018} for details. 
Therefore, without any additional assumptions on the structure of the underlying 
time series, $p$ can only be as large as $o(n^{1/2})$. 
To deal with the case of large $p$, we impose some conditions on the transformation 
matrix $\bL$ of Equation (\ref{decom}) and the cross dependence of time series $\by_t$. 

Let $\bL=[\bfc_1,...,\bfc_p]$, where $\bfc_i$ is a $p$-dimensional column vector, and $\bL_1=[\bfc_1,...,\bfc_r]$ and $\bL_2=[\bfc_{r+1},...,\bfc_p]$.

\begin{assumption}
(i) $\bL_1=[\bfc_1,...,\bfc_r]$ such that $\|\bfc_j\|_2^2\asymp p^{1-\delta_1}$, $j=1,...,r$ and $\delta_1\in[0,1)$; (ii) For each $j=1,...,r$ and $\delta_1$ given in (i), $\min_{\theta_i\in R,i\neq j}\|\bfc_j-\sum_{1\leq i\leq r,i\neq j}\theta_i\bfc_i\|_2^2\asymp p^{1-\delta_1}$.
\end{assumption}

\begin{assumption}
(i) $\bL_2$ admits a singular value decomposition $\bL_2=\bA_2\bD_2\bV_2'$, where $\bA_2\in R^{p\times v}$ is given in Equation (\ref{sfactor}), $\bD_2=\diag(d_1,...,d_v)$ and $\bV_2\in R^{v\times v}$ satisfying $\bV_2'\bV_2=\bI_v$; (ii) There exists a finite integer $0< K< v$ such that $d_1\asymp...\asymp d_K\asymp p^{(1-\delta_2)/2}$ for some $\delta_2\in[0,1)$ and $d_{K+1}\asymp...\asymp d_v\asymp 1$.
\end{assumption}


\begin{assumption}
(i) For $\gamma$ given in Assumption 1, any $\bh\in R^{v}$ and $0<c_h<\infty$ with $\|\bh\|_2=c_h$, $E|\bh'\bve_t|^{2\gamma}<\infty$; (ii)  $\sigma_{\min}(\bR'\bB_2^{*}{'}\bA_1)\geq C_3$ for some constant $C_3>0$ and some half orthogonal matrix $\bR\in R^{(p-K)\times r}$ satisfying $\bR'\bR=\bI_r$, where $\sigma_{\min}$ denotes the minimum non-zero singular value of a matrix. 
\end{assumption}

The quantity $\delta_1$ of Assumption 4 is used to quantify the strength of the factors, and the eigenvalues of $\bL_1\bL_1'$ are all of order $p^{1-\delta_1}$.
If $\delta_1=0$, the corresponding factors are called strong factors, since it includes the case where each element of $\bfc_i$ is $O(1)$. If $\delta_1>0$, the corresponding 
factors are weak factors and the smaller the $\delta_1$ is, the stronger the factors are. 
An advantage of using index $\delta_1$ is to link the convergence rates of the estimated factors explicitly to the strength of the factors. Assumption 4 ensures that all common factors in $\bx_t$ are of equal strength $\delta_1$. In practice, the factors may have multiple levels of  strength as in \cite{lamyao2012}, among others. We can make similar assumptions and the consistency of the loading matrix would then depend on the strength of the weakest factors. We do not consider this issue here to save space. 
There are many sufficient conditions  for Assumption 5 to hold. For example, it holds if we allow $(\bfc_{r+1},...,\bfc_{r+K})$ to satisfy Assumption 4 for some $\delta_2\in[0,1)$, and the $L_1$- and $L_\infty$-norms of $(\bfc_{r+K+1},...,\bfc_p)$ are all finite. A special case is to let $\bfc_{r+K+j}$ be a standard unit vector. The constraint between $\delta_1$ and $\delta_2$ will be illustrated later under different scenarios to guarantee the consistency. Assumption 6(i) is mild and includes the  standard normal distribution as a special case. Together with Assumption 2 and the aforementioned sufficient condition for Assumption 5, it is not hard to show $E|y_{it}|^{2\gamma}<\infty$, but we do not address this issue explicitly here. Assumption 6(ii) is reasonable since $\bB_2$ is a subspace of $\bB_2^*$, and Remark 1 implies that $\bR'\bB_2^{*}{'}\bA_1$ is invertible. The choice of $\wh\bR$ and hence $\wh\bB_2=\wh{\bB}_2^*\wh\bR$ will be discussed later.

\begin{remark}
In Assumption 5, we actually only require $d_K\asymp p^{(1-\delta_2)/2}$ for some $\delta_2\in[0,1)$ and $K\geq 1$, and the upper singular values $\{d_1,...,d_{K-1}\}$ if $K>1$ can be even larger provided that the largest one $d_1$ should be bounded by another rate $p^{(1-\delta_3)/2}$ for some $0\leq \delta_3\leq \delta_2$. 
For simplicity, we assume the top singular values are of the same order. 
\end{remark}
If the dimension $p$ is large, it is not possible to consistently estimate $\bB_2$ or even $\mathcal{M}(\bB_2)$. Instead, we estimate $\bB_2^*=(\bA_{22},\bB_2)$ or equivalently $\mathcal{M}(\bB_2^*)$, which is the subspace spanned by the eigenvectors associated with the $p-K$ smallest eigenvalues of $\bS$. Assume $\wh{\bB}_2^*$ consists of the eigenvectors corresponding to the smallest $p-K$ eigenvalues of $\wh\bS$. Under some conditions, we can show that $\mathcal{M}(\wh{\bB}_2^*)$ is consistent to $\mathcal{M}(\bB_2^*)$. This is also the case in the literature on high-dimensional PCA with i.i.d. data. See, for example, \cite{shenetal2016} and the references therein. Therefore, the choice of $\wh\bB_2$ should be a subspace of $\wh{\bB}_2^*$, and we will discuss it before Theorem \ref{tm4} below. If there exist cross-correlations between $\bff_t$ and $\bve_{t-j}$ for $j\geq 1$, we assume rank($\bSigma_{f\epsilon}(k)$)$=r$ and define
\begin{equation}\label{kmm}
\kappa_{\min}=\min_{1\leq k\leq k_0}\|\bSigma_{f\epsilon}(k)\|_{\min}\,\, \text{and}\,\, \kappa_{\max}=\max_{1\leq k\leq k_0}\|\bSigma_{f\epsilon}(k)\|_2,
\end{equation}
where $\|\cdot\|_{\min}$ denotes the smallest nonzero singular value, $\kappa_{\min}$ and $\kappa_{\max}$ can be either  finite constants or diverging rates in relation to $p$ and $n$, and they control the strength of the dependence between $\bff_t$ and the past errors $\bve_{t-j}$ for $j\geq 1$. The maximal order of $\kappa_{\max}$ is $p^{1/2}$ which is the Frobenius norm of $\bSigma_{f\ve}(k)$ and $\kappa_{\max}=0$ (hence $\kappa_{\min}=0$) implies that $\bff_t$ and $\bve_s$ are independent for all $t$ and $s$.
Throughout this article, if $\bff_t$ and $\bve_s$ are independent for all $t$ and $s$, then $\kappa_{\min}=\kappa_{\max}=0$ and all the conditions and expressions below concerning  $\kappa_{\min}$ and $\kappa_{\max}$ are removed.

\begin{theorem}\label{tm2}
Suppose Assumptions 1-6 hold and $r$ is known and fixed. As $n\rightarrow\infty$, if $p^{\delta_1}n^{-1/2}=o(1)$ or $\kappa_{\max}^{-1}p^{\delta_1/2+\delta_2/2}n^{-1/2}=o(1)$, then
\[D(\wh\bA_1,\bA_1)=\left\{\begin{array}{ll}
O_p(p^{\delta_1}n^{-1/2}),&\text{if}\,\, \kappa_{\max}p^{\delta_1/2-\delta_2/2}=o(1),\\
O_p(\kappa_{\min}^{-2}p^{\delta_2}n^{-1/2}+\kappa_{\min}^{-2}\kappa_{\max}p^{\delta_1/2+\delta_2/2}n^{-1/2}),&\text{if}\,\, r\leq K, \kappa_{\min}^{-1}p^{\delta_2/2-\delta_1/2}=o(1),\\
O_p(\kappa_{\min}^{-2}pn^{-1/2}+\kappa_{\min}^{-2}\kappa_{\max}p^{1+\delta_1/2-\delta_2/2}n^{-1/2}),&\text{if}\,\, r>K, \kappa_{\min}^{-1}p^{(1-\delta_1)/2}=o(1),
\end{array}\right.\]
and the above results also hold for $D(\wh\bB_1,\bB_1)$.
Furthermore,
\[ D(\wh{\bB}_2^*,\bB_2^*)=O_p\left(p^{2\delta_2-\delta_1}n^{-1/2}+p^{\delta_2}n^{-1/2}+(1+p^{2\delta_2-2\delta_1})D(\wh\bB_1,\bB_1)\right).\]

\end{theorem}

\begin{remark}
(i) If $\kappa_{max}=\kappa_{\min}=0$, i.e., $\bff_t$ and $\bve_s$ are independent for all $t$ and $s$, we have
\[D(\wh\bA_1,\bA_1)=O_p(p^{\delta_1}n^{-1/2})\,\,\text{and}\,\,D(\wh{\bB}_2^*,\bB_2^*)=O_p(p^{2\delta_2-\delta_1}n^{-1/2}+p^{\delta_2}n^{-1/2}+p^{\delta_1}n^{-1/2}).\]
To guarantee that these estimates are consistent, we require $p^{\delta_1}n^{-1/2}=o(1)$, $p^{\delta_2}n^{-1/2}=o(1)$ and $p^{2\delta_2-\delta_1}n^{-1/2}=o(1)$. When $p\asymp n^{1/2}$, it implies that $0\leq \delta_1< 1$, $0\leq \delta_2<1$ and $\delta_2<(1+\delta_1)/2$, i.e., the ranges of $\delta_1$ and $\delta_2$ are pretty wide. On the other hand, if $p\asymp n$, we see that $0\leq \delta_1< 1/2$, $0\leq \delta_2<1/2$ and $2\delta_2-\delta_1<1/2$, these ranges become narrower if $p$ is large.\\
(ii) When $\kappa_{\max}\neq 0$ and $\kappa_{\min}\neq 0$, there are many possible results. A reasonable assumption is $\kappa_{\min}\asymp\kappa_{\max}\asymp p^{\delta/2}$ for some $0\leq\delta< 1$ since $r$ is small. For example, set $\delta=\delta_1$, 
\[D(\wh\bA_1,\bA_1)=\left\{\begin{array}{ll}
O_p(p^{\delta_1}n^{-1/2}),&\text{if}\,\,p^{\delta_1-\delta_2/2}=o(1),\\
O_p(p^{\delta_2/2}n^{-1/2}),&\text{if}\,\, r\leq K, \kappa_{\min}^{-1}p^{\delta_2/2-\delta_1/2}=o(1),
\end{array}\right.\]
and there is no consistency result when $r>K$. Furthermore, we have $D(\wh{\bB}_2^*,\bB_2^*)=O_p(p^{2\delta_2-\delta_1}n^{-1/2} +p^{\delta_2}n^{-1/2})$. Thus, we require $p^{\delta_1}n^{-1/2}=o(1)$, $p^{\delta_2}n^{-1/2}=o(1)$ and $p^{2\delta_2-\delta_1}n^{-1/2}$ $=o(1)$. The ranges of $\delta_1$ and $\delta_2$ are the same as discussed in Remark 3(i) above, we omit the details here. On the other hand, if $\delta>(1-\delta_1)/2$, it is still possible to obtain consistent estimates when $r>K$, the discussion is similar and is omitted for simplicity. 
\end{remark}

From Theorem \ref{tm2}, we see that when $p\asymp n$, we  require $\delta_1<1/2$ and $\delta_2<1/2$ to guarantee the consistency of our estimation method, which rules out the cases 
of the presence of weaker factors with $\delta_1\geq 1/2$ and a slower diverging of the noise covariance matrix with $ \delta_2\geq 1/2$. The convergence rates in Theorem \ref{tm2} are not optimal and they can be further improved under some additional assumption on $\bve_t$ below.
\begin{assumption}
For any $\bh\in R^{v}$ with $\|\bh\|_2=1$, there exists a constant $C_4>0$ such that
\[P(|\bh'\bve_t|>x)\leq 2\exp(-C_4x)\quad \text{for any}\,\, x>0.\]
\end{assumption}
Assumption 7 implies that $\bve_{t}$ are sub-exponential, which is a larger class of distributions than sub-gaussian, and  includes the uniform distribution on every convex body following the Brunn-Minkowski inequality. See, for example, \cite{vershynin2012} and \cite{vershynin2018}.
\begin{theorem}\label{tm3}
Suppose Assumptions 1-7 hold and $r$ is known and fixed, and $p^{\delta_1/2}n^{-1/2}=o(1)$, $p^{\delta_2/2}n^{-1/2}=o(1)$. 
(i) Under the condition that $\delta_1\leq \delta_2$,
\[D(\wh\bA_1,\bA_1)=\left\{\begin{array}{ll}
O_p(p^{\delta_1/2}n^{-1/2}),&\text{if}\,\, \kappa_{\max}p^{\delta_1/2-\delta_2/2}=o(1),\\
O_p(\kappa_{\min}^{-2}p^{\delta_2-\delta_1/2}n^{-1/2}+\kappa_{\min}^{-2}\kappa_{\max}p^{\delta_2/2}n^{-1/2}),&\text{if}\,\, r\leq K, \kappa_{\min}^{-1}p^{\delta_2/2-\delta_1/2}=o(1),\\
O_p(\kappa_{\min}^{-2}p^{1-\delta_1/2}n^{-1/2}+\kappa_{\min}^{-2}\kappa_{\max}p^{1-\delta_2/2}n^{-1/2}),&\text{if}\,\, r>K, \kappa_{\min}^{-1}p^{(1-\delta_1)/2}=o(1),
\end{array}\right.\]
and the above results also hold for $D(\wh\bB_1,\bB_1)$, and 
\[D(\wh{\bB}_2^*,\bB_2^*)=O_p(p^{2\delta_2-3\delta_1/2}n^{-1/2}+p^{2\delta_2-2\delta_1}D(\wh\bB_1,\bB_1)).\]

(ii) Under the condition that $\delta_1>\delta_2$, if $\kappa_{\max}=0$ and $p^{\delta_1-\delta_2/2}n^{-1/2}=o(1)$, then
\[D(\wh\bA_1,\bA_1)=O_p(p^{\delta_1-\delta_2/2}n^{-1/2}).\]
If $\kappa_{\max}>>0$,
then
\[D(\wh\bA_1,\bA_1)=\left\{\begin{array}{ll}
O_p(\kappa_{\min}^{-2}\kappa_{\max}p^{\delta_1/2}n^{-1/2}),&\text{if}\,\, r\leq K, \kappa_{\min}^{-1}p^{\delta_2/2-\delta_1/2}=o(1),\\
O_p(\kappa_{\min}^{-2}\kappa_{\max}p^{1+\delta_1/2-\delta_2}n^{-1/2}),&\text{if}\,\, r>K, \kappa_{\min}^{-1}p^{(1-\delta_1)/2}=o(1),
\end{array}\right.\]
and the above results also hold for $D(\wh\bB_1,\bB_1)$, and
\[D(\wh{\bB}_2^*,\bB_2^*)=O_p(p^{\delta_2/2}n^{-1/2}+D(\wh\bB_1,\bB_1)).\]

\end{theorem}


\begin{remark}
(i) Consider the case $\kappa_{\min}=\kappa_{\max}=0$. If $\delta_1\leq \delta_2$, $D(\wh\bA_1,\bA_1)=O_p(p^{\delta_1/2}n^{-1/2})$ and $D(\wh{\bB}_2^*,\bB_2^*)=O_p(p^{2\delta_2-3\delta_1/2}n^{-1/2})$. For $p\asymp n$, we require $0\leq \delta_1\leq \delta_2<1$ and $4\delta_2-3\delta_1<1$, or equivalently $0\leq \delta_1\leq \delta_2<3\delta_1/4+1/4$. If $\delta_1>\delta_2$, $D(\wh\bA_1,\bA_1)=O_p(p^{\delta_1-\delta_2/2}n^{-1/2})$ and $D(\wh{\bB}_2^*,\bB_2^*)=O_p(p^{\delta_1-\delta_2/2}n^{-1/2})$. Thus, if $p\asymp n$, we require $\max\{2\delta_1-1,0\}<\delta_2<\delta_1<1$. Therefore, if $\bff_t$ and $\bve_s$ are 
independent and $p\asymp n$, $\delta_1$ and $\delta_2$ need to satisfy $0\leq \delta_1\leq \delta_2<3\delta_1/4+1/4$ or $\max\{2\delta_1-1,0\}<\delta_2<\delta_1<1$, 
which is much wider than those of Theorem 3.\\
(ii) If $\bff_t$ and $\bve_s$ are correlated for $s<t$, we may have many consistency results depending on the strength of the dependence between $\bff_t$ and $\bve_s$. In addition, if rank$(\bSigma_{f\ve})<r$, the conditions $r\leq K$ and $r>K$ in Theorems \ref{tm2} and \ref{tm3} would become rank$(\bSigma_{f\ve})\leq K$ and rank$(\bSigma_{f\ve})> K$, respectively. 
We omit the details here.\\
(iii) When $K$ increases with $p$ and $n$, similar results can still be established, depending on the growing rates of $K$, $p$, and $n$. See also Remark 2(c) in \cite{fan2013}. But we do not pursue this issue here.
\end{remark}

By Remark 1 and Assumption 6, so long as $\mathcal{M}(\wh\bA_1)$ and $\mathcal{M}(\wh\bB_2^*)$ are consistent estimators for $\mathcal{M}(\bA_1)$ and $\mathcal{M}(\bB_2^*)$, respectively, there must exist an $\bR$ which is the same as that in Assumption 6(ii) such that $\mathcal{M}(\wh\bB_2^*\bR$) is a consistent estimator for $\mathcal{M}(\bB_2)$, which implies the invertibility of $\bR'\wh\bB_2^*{'}\wh\bA_1$. Therefore, once having $\wh{\bB}_2^*$, we suggest to choose $\wh\bB_2$ as $\wh\bB_2=\wh{\bB}_2^*\wh\bR$, 
where $\wh\bR=(\wh\br_1,...,\wh\br_r)\in R^{(p-K)\times r}$, and $\wh\br_i$ is the vector associated with the $i$-th largest eigenvalues of $\wh{\bB}_2^*{'}\wh\bA_1\wh\bA_1'\wh{\bB}_2^*$. This choice guarantees that the matrix $(\wh\bB_2'\wh\bA_1)^{-1}$ behaves well when recovering the factor $\wh\bx_t$ and is sufficient for our analysis. On the other hand, this choice could still eliminate the diverging part of the noise covariance matrix and gives prominent convergence rate, as shown in Theorem \ref{tm4}.  There are many ways to choose the number of components $K$ in Assumption 5 so long as $p-K>r$. We will discuss the choice of $K$ in Remark 5 below and also in Section 5. The following theorem states the convergence rate of the extracted common factors. 


\begin{theorem}\label{tm4}
Under the conditions in Theorem \ref{tm2} or \ref{tm3}, we have
\begin{equation}\label{ex:f}
p^{-1/2}\|\wh\bA_1\wh\bx_t-\bA_1\bx_t\|_2=O_p\{p^{-1/2}+p^{-\delta_1/2}D(\wh\bA_1,\bA_1)+p^{-\delta_2/2}D(\wh{\bB}_2^*,\bB_2^*)\}.
\end{equation}
Furthermore, if $\bff_t$ satisfies Assumption 7 with a constant $C_5>0$, we have
\[p^{-1/2}\max_{1\leq t\leq n}\|\wh\bA_1\wh\bx_t-\bA_1\bx_t\|_2=O_p\{p^{-1/2}\log(n)+p^{-\delta_1/2}\log(n)D(\wh\bA_1,\bA_1)+p^{-\delta_2/2}\log(n)D(\wh{\bB}_2^*,\bB_2^*)\}.\]
\end{theorem}


\begin{remark}
(i) A similar result is given in Theorem 3 of \cite{LamYaoBathia_Biometrika_2011} for the approximate factor models. The above results also hold for $\|\wh{\bA_2\be_t}-\bA_2\be_t\|_2$ by a simple manipulation, where $\wh{\bA_2\be_t}=\by_t-\wh\bA_1\wh\bx_t$. In addition, if all the eigenvalues associated with the idiosyncratic covariance are bounded as in \cite{LamYaoBathia_Biometrika_2011}, by the proof of Theorem 5 in the supplement, $p^{-\delta_2/2}D(\wh{\bB}_2^*,\bB_2^*)$ will disappear and (\ref{ex:f}) reduces to Theorem 3 in \cite{LamYaoBathia_Biometrika_2011}, and we can still consistently estimate the extracted factors. When  $\delta_1=\delta_2=0$, i.e., the factors and the noise terms are all strong, the convergence rate in (\ref{ex:f}) is $O_p(p^{-1/2}+n^{-1/2})$, which is the optimal rate specified in Theorem 3 of \cite{Bai_Econometrica_2003} when dealing with the traditional approximate factor models.

(ii) Selecting the number of principal components is a common issue in the literature,  and  many approaches are available. 
Since it is impossible to eliminate all the noise effects in recovering the factors 
and we only need to guarantee that the diverging part of the noises 
are removed for large $p$, we may select $K$ in a range of possible values. 
In practice, let $\wh\mu_1\geq...\geq  \wh\mu_p$ be the sample eigenvalues of $\wh\bS$ and define $\wh K_L$ as
\begin{equation}\label{kl}
 \wh K_L=\arg\min_{1\leq j\leq \wh K_U}\{\wh\mu_{j+1}/\wh\mu_{j}\}.
\end{equation} 
Let $\wh K_U$ be a pre-specified integer. In practice, we suggest $\wh K_U=\min\{\sqrt{p},\sqrt{n},p-\wh r,10\}$. Then the estimator $\wh K$ of $K$ can assume some value between $\wh K_L$ and $\wh K_U$. 
\end{remark}

Next, we study the consistency of the white noise test described in Section 2. The consistency conditions depend on the test statistic used. In what follows, 
we only present the consistency for large $p$ since the case of small $p$ is trivial.

\begin{theorem}\label{tm5}
(i) Assume that Assumptions 1-7 hold. If $D(\wh\bB_1,\bB_1)^2\|\bSigma_y\|_2=o_p(1)$, then the test statistic $T_n$ defined in (\ref{t:test}) can consistently estimate $r$, i.e., $P(\wh r=r)\rightarrow 1$ as $n\rightarrow\infty$.\\
(ii) Suppose Assumptions 1-7 hold and $\bff_t$ also satisfies Assumption 7. If \[p^{1/2}\log(np)\max\{p^{\delta_2/2-\delta_1/2}D(\wh\bB_1,\bB_1),D(\wh\bB_1,\bB_1),O_p(p^{\delta_2/2}n^{-1/2})\}=o_p(1),\] then 
the test statistic $T(m)$ of  (\ref{Tsay:test}) can also consistently estimate $r$.
\end{theorem}


\begin{remark}
(i) If $\bff_t$ and $\bve_s$ are independent for all $t$ and $s$, the conditions in Theorem \ref{tm5}(i) are essentially $pn^{-1}=o(1)$ if $\delta_1\leq \delta_2$ and $p^{1+2\delta_1-2\delta_2}n^{-1}=o(1)$ if $\delta_1>\delta_2$. Thus, we require $p\asymp n^{\xi}$ with $0<\xi<1$ for both cases. As for Theorem \ref{tm5}(ii), the condition is 
$p^{(1+\delta_2)/2}n^{-1/2}\log(np)=o(1)$ if $\delta_1\leq \delta_2$ and $p^{1/2+\delta_1-\delta_2/2}n^{-1/2}\log(np)=o(1)$ if $\delta_1>\delta_2$, and hence we also require $p\asymp n^{\xi}$ for some $0<\xi<1$.\\
(ii) Even though the conditions in Theorem 5(ii) are slightly stronger, 
 the method based on $T(m)$ is simple and easy to use, and its performance is satisfactory when $p$ is moderately large. See \cite{Tsay_2018} and the simulation results in Section 4 for details. 
We use those stronger conditions, 
because we require the effect of the orthogonalization using sample PCA  on the rank of each component of the white noise part is asymptotically negligible uniformly in terms of $t$ and $p$ in 
the proofs. 
\end{remark}

With the estimator $\wh r$, we may define the estimator for $\bA_1$ as $\wt\bA_1=(\wh\ba_1,...,\wh\ba_{\wh r})$, where $\wh\ba_1,...,\wh\ba_{\wh r}$ are the orthonormal eigenvectors of $\wh \bM$, defined in (\ref{mhat}), corresponding to the $\wh r$ largest eigenvalues. In addition, we may also replace $r$ by $\wh r$ in the methodology described in Section 2. To see the impact of $\wh r$ on measuring the errors in estimating the loading spaces, we take $\wt \bA_1$ for example and use
\begin{equation}\label{D2}
\wt D(\wt\bA_1,\bA_1)=\sqrt{1-\frac{1}{\max(\wh r,r)}\tr(\wt\bA_1\wt\bA_1'\bA_1\bA_1')},
\end{equation}
which is a modified version of (\ref{eq:D}).  It takes into account the fact that the dimensions of $\wt\bA_1$ and $\bA_1$ may be different, and we use $\max(r_1,r_2)$ instead of $\min(r_1,r_2)$ to guarantee that the quantity in the square-root is nonnegative.  We also see that $\wt D(\wt\bA_1,\bA_1)=D(\wh\bA_1,\bA_1)$ if $\wh r=r$. We show below that $\wt D(\wt\bA_1,\bA_1)\rightarrow 0$ in probability at the same rate of $D(\wh\bA_1,\bA_1)$. Therefore, $\mathcal{M}(\wt\bA_1)$ is still a consistent estimator for $\mathcal{M}(\bA_1)$ even without knowing $r$. Let $\rho_n$ be the convergence rate of $D(\wh\bA_1,\bA_1)$ as in Theorem 1, 2, or 3, that is, $\rho_nD(\wh\bA_1,\bA_1)=O_p(1)$. For any $\epsilon>0$, there exists a positive constant $M_{\epsilon}$ such that $P(\rho_nD(\wh\bA_1,\bA_1)>M_{\epsilon})<\epsilon$. Then,
\begin{align*}
P(\rho_n\wt D(\wt\bA_1,\bA_1)>M_{\epsilon})&\leq P(\rho_n D(\wh\bA_1,\bA_1)>M_{\epsilon},\wh r=r)+P(\rho_n\wt D(\wt\bA_1,\bA_1)>M_{\epsilon},\wh r\neq r)\\
&\leq P(\rho_n D(\wh\bA_1,\bA_1)>M_{\epsilon},\wh r=r)+o(1)\leq \epsilon+o(1)\rightarrow \epsilon,
\end{align*}
which implies $\rho_n\wt D(\wt\bA_1,\bA_1)=O_p(1)$. Therefore, $\mathcal{M}(\wt\bA_1)$ has the oracle property in estimating the factor loading space $\mathcal{M}(\bA_1)$ in the sense that it has the same convergence rate as $D(\wh\bA_1,\bA_1)$. The result also holds for $\wh\bB_1$ by a similar argument. Therefore, we still use $\wh\bA_1$ and $\wh\bB_1$ as our estimators when replacing $r$ by $\wh r$. As for the influence of  $\wh r$ and $\wh K$ on $\wh\bB_2$, note that we do not have the parameter $K$ when $p$ is small, and hence the impact of $\wh r$ on $\wh\bB_2$ remains  negligible asymptotically by a similar argument. When $p$ is large, note that we only eliminate the possible diverging effects of the noises in our approach. Thus, an accurate estimator for $K$ is not necessary and any estimator $\wh K$ which is larger than the true one is sufficient for our method to work. On the other hand, we can show  $\wh K_L$ is a consistent estimator for $K$ under certain conditions by a similar argument as that of \cite{lamyao2012}. Therefore, we may set an upper bound $\wh K_U$ as in Remark 5(ii) and choose $\wh K$ between $\wh K_L$ and $\wh K_U$. The choice of $\wh K_U$ is not unique but it cannot exceed $p-\wh r$ since the rank of $\wh\bS$ is at most $p-\wh r$. The optimal one can be chosen via some cross-validation such as the 
out-of-sample testing if one is concerned with the forecasting performance of the extracted factors. 
For further information, see the empirical example in Section 4 and the concluding remark 
in Section 5.


\section{Numerical Properties}

We use simulation and a real example to assess the performance of the proposed 
analysis in finite samples. An additional real example is given in the online supplement. 

\subsection{Simulation}
In this section, we illustrate the finite-sample properties of the proposed methodology under the scenarios when $p$ is both small and large. 
As the dimensions of $\wh\bA_1$  and $\bA_1$ are not necessarily the same,
and  $\bL_1$ is not an orthogonal matrix in general, we first extend the discrepancy measure
in Equation (\ref{D2}) to a more general one below. Let $\bH_i$ be a
$p\times r_i$ matrix with rank$(\bH_i) = r_i$, and $\bP_i =
\bH_i(\bH_i'\bH_i)^{-1} \bH_i'$, $i=1,2$. Define
\begin{equation}\label{dmeasure}
\bar{D}(\bH_1,\bH_2)=\sqrt{1-
\frac{1}{\max{(r_1,r_2)}}\textrm{tr}(\bP_1\bP_2)},
\end{equation}
then $\bar{D} \in [0,1]$. Furthermore,
$\bar{D}(\bH_1,\bH_2)=0$ if and only if
either $\mathcal{M}(\bH_1)\subset \mathcal{M}(\bH_2)$ or
$\mathcal{M}(\bH_2)\subset \mathcal{M}(\bH_1)$, and  it is 1 if and only if
$\mathcal{M}(\bH_1) \perp \mathcal{M}(\bH_2)$.
When $r_1 = r_2=r$ and $\bH_i'\bH_i= \bI_r$,
$\bar{D}(\bH_1,\bH_2)$ is the same as that in Equation (\ref{eq:D}). 
We only present the simulation results 
for $k_0=2$ in Equation (\ref{mhat}) to save space because other choices of $k_0$ 
produce similar patterns as shown in the online supplement.\\

{\noindent \bf Example 1.} Consider Model (\ref{decom}) with common factors following the VAR(1) model 
\[\bff_t=\bPhi\bff_{t-1}+\bfeta_t,\]
where $\bfeta_t$ is a white noise process. 
 We set the true number of factors $r=3$, the dimension $p= 5, 10, 15, 20$, and the sample size $n=200, 500, 1000, 1500, 3000$. For each realization of $\by_t$, 
 the elements of the loading matrix $\bL$ are drawn independently from $U(-2,2)$, and the elements of $\bL_2$ are then divided by $\sqrt{p}$ to balance the accumulated variances of $f_{it}$ and $\ve_{it}$ for each component of $\by_t$. $\bPhi$ is a diagonal matrix with its diagonal elements being drawn independently from $U(0.5,0.9)$, $\bve_t\sim N(0,\bI_v)$ and $\bfeta_t\sim N(0,\bI_r)$. We use 1000 replications for each $(p,n)$ configuration.
 
\begin{figure}
\begin{center}
\subfigure[]{\includegraphics[width=0.50\textwidth]{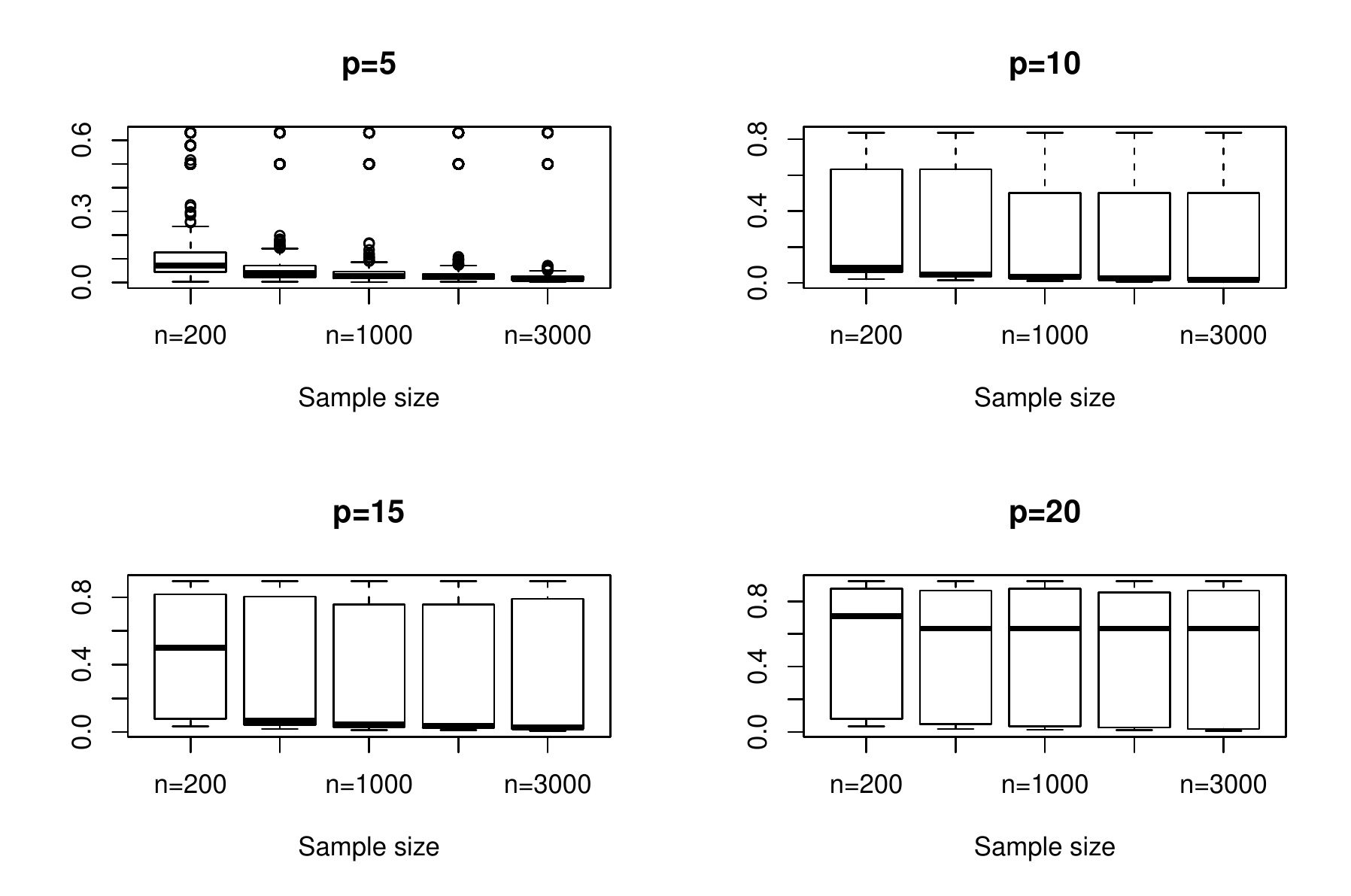}}
\subfigure[]{\includegraphics[width=0.45\textwidth]{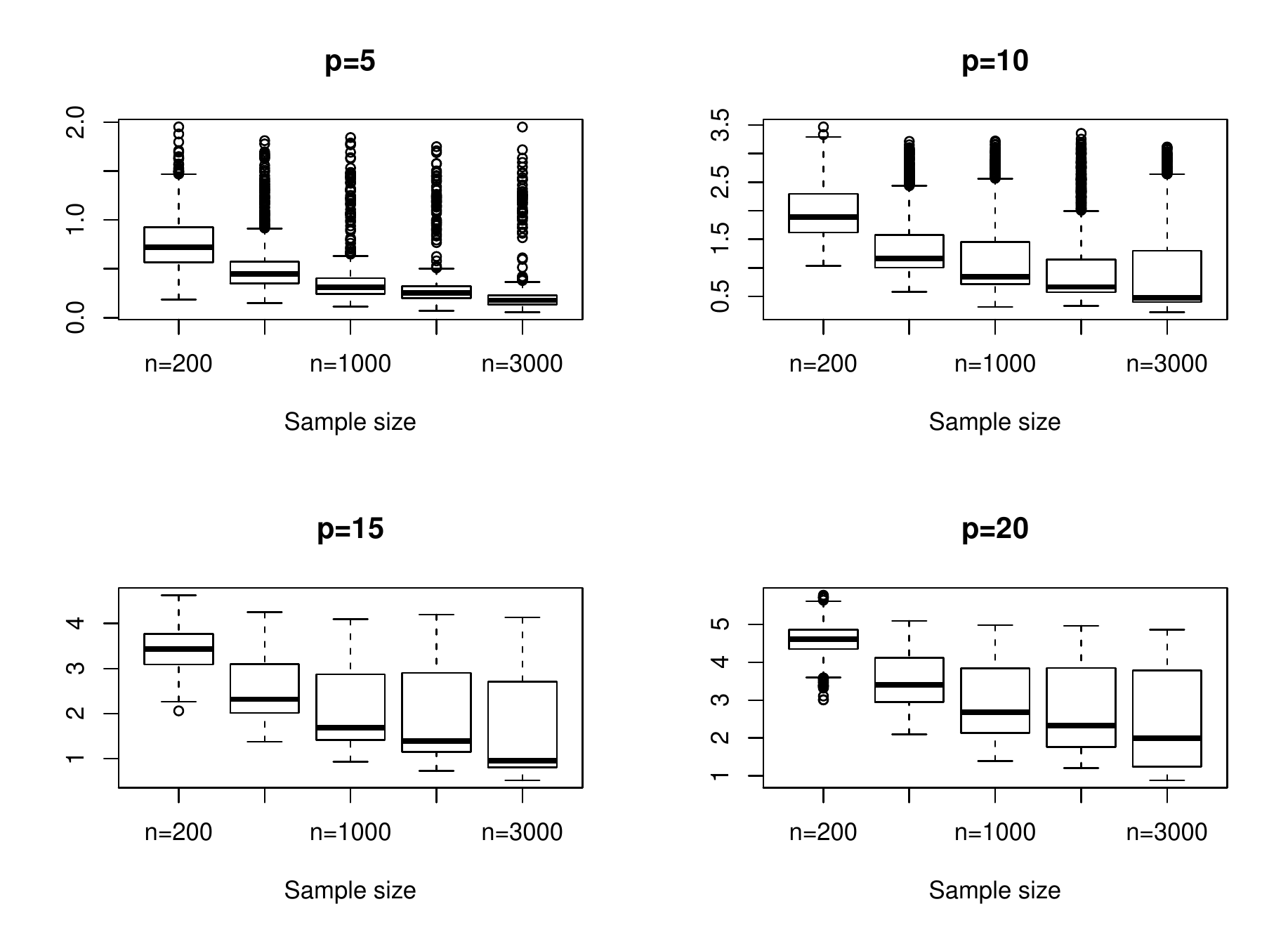}}
\caption{(a) Boxplots of $\bar{D}(\wh\bA_1,\bL_1)$ when $r=3$ under the scenario that $p$ is relatively small in Example 1; (b) Boxplots of the RMSE defined in (\ref{rmse:psm}) when $r=3$ under the scenario that $p$ is relatively small in Example 1. The sample sizes used are $200, 500, 1000, 1500, 3000$, and the results are based on $1000$ iterations.}\label{fig1}
\end{center}
\end{figure}

We first study the performance of estimating the number of factors. Since $p$ is relatively small compared to the sample size $n$, we use Ljung-Box test statistics with 
$m$ = 10 to determine the number of factors, i.e., $Q(10)$. The empirical probabilities $P(\wh r=r)$ are reported in Table~\ref{Table1}. From the table, we see that, for a given $p$, the performance of the proposed method improves as the sample size increases. On the other hand, for a given $n$, the empirical probability decreases as $p$ increases and the probability for $p=20$ is roughly half of that for $p=5$, which is understandable since it is harder to determine the correct number of factors when the dimension increases and the errors in the testing procedure accumulate. 
Overall, the Ljung-Box test works well for the 
case of small dimension (e.g., $p\leq 10$). However, when $p$ is slightly larger (e.g., $p=15, 20$), the test statistic tends to overestimate the number of factors, implying that we can still keep 
sufficient information of the original process $\by_t$. To illustrate, we present the boxplots of $\bar{D}(\wh\bA_1,\bL_1)$ in Figure~\ref{fig1}(a), where $\bar{D}(\cdot,\cdot)$ is defined in (\ref{dmeasure}). From Figure~\ref{fig1}(a), for each $p$, the discrepancy decreases as the sample size increases and this is in agreement with the theory. The plot also shows that, as expected, 
the mean discrepancy increases as the dimension $p$ increases.

Furthermore, for each $(p,n)$, we study the root-mean-square error (RMSE):
\begin{equation}\label{rmse:psm}
\text{RMSE}=\left(\frac{1}{n}\sum_{t=1}^n\|\wh\bA_1\wh\bx_t-\bL_1\bff_t\|_2^2\right)^{1/2},
\end{equation} 
which quantifies the accuracy in estimating the common factor process.  
Boxplots of the RMSE are shown in Figure~\ref{fig1}(b). From the plot, we  see a clear pattern that, as the sample size increases, the RMSE decreases for a given $p$, which is consistent with the results of Theorem 1. Obviously, as expected, the RMSE increases with the dimension $p$. 
Overall, the one-by-one testing procedure for determining $r$ 
works well when the dimension of 
$\by_t$ is small, and the RMSE  decreases when the sample size increases, 
but the performance of the test may deteriorate into overestimation of the number of the factors for larger  $p$.

\begin{table}
 \caption{Empirical probabilities $P(\hat{r}=r)$ of various $(p,n)$ configurations 
 for the model of Example 1 with $r=3$, where $p$ and $n$ are the dimension and the sample size, respectively. $1000$ iterations are used.} 
          \label{Table1}
\begin{center}
 \setlength{\abovecaptionskip}{0pt}
\setlength{\belowcaptionskip}{3pt}

\begin{tabular}{c|c|ccccc}
\hline
&&\multicolumn{5}{c}{$n$}\\
&$p$&$200$&$500$&$1000$&$1500$&$3000$\\
\hline
$r=3$&$5$&0.861&0.889&0.890&0.912&0.926\\
&$10$&0.683&0.718&0.723&0.735&0.748\\
&$15$&0.506&0.555&0.561&0.599&0.601\\
&$20$&0.395&0.425&0.441&0.447&0.453\\
   \hline 
\end{tabular}
  \end{center}
\end{table}


{\noindent \bf Example 2.} In this example, we consider Model (\ref{decom}) with $\bff_t$ 
being the same as that of Example 1. We set the true number of factors  $r=5$ and the number of the spiked components  $K$ = 3 and 7 defined in Assumption 5 for the loading of white noise processes. 
The dimensions used are $p=50, 100, 300, 500$, and the sample sizes are $n=300, 500, 1000, 1500, 3000$. We consider three scenarios for the strength parameters $\delta_1$ and $\delta_2$: $(\delta_1,\delta_2)=(0,0)$, $(\delta_1,\delta_2)=(0.4,0.5)$, and $(\delta_1,\delta_2)=(0.5,0.4)$. For each realization of $\by_t$, 
the elements of the loading matrix 
$\bL$ are drawn independently from $U(-2,2)$, and then we divide $\bL_1$ by $p^{\delta_1/2}$,  the first $K$ columns of $\bL_2$ by $p^{\delta_2/2}$, and the remaining $v-K$ columns by $p$ to satisfy Assumptions 4 and 5. $\bPhi$, $\bve_t$ and $\bfeta_t$ are drawn similarly as those of Example 1. We also use $1000$ replications in each experiment.

We first study the performance of the high-dimensional white noise test. For simplicity, we only present the results of the $T(m)$ statistics defined in (\ref{Tsay:test}) and 
the results for the other test are similar. When $p\geq n$, we only keep the upper $\epsilon n$ components of $\wh\bG'\by_t$ with $\epsilon=0.75$ in the testing and the test still works well for other choices of $\epsilon$ as shown in the online supplement. The testing results are reported in Table~\ref{Table2} for $r=5$ with $K=3$ and $K=7$, respectively. From Table~\ref{Table2}, we see that for each setting of $(\delta_1,\delta_2)$ and fixed $p$, the performance of the white noise test improves as the sample size increases. 
The performance is also quite satisfactory for moderately larger $p$. Also, for $n \geq 1000$, 
the empirical probabilities of selecting the correct number of factors are high for both $K=3$ and 
$K=7$. 

To shed some light on the advantages of the proposed methodology, we compare it 
with that  of \cite{LamYaoBathia_Biometrika_2011} (denoted by LYB) 
 in selecting the number of factors. 
For the ratio-based method in LYB, let $\wh \lambda_1,...,\wh\lambda_p$ be the eigenvalues of $\wh\bM$ and define
\begin{equation}\label{ratio:lby}
\wh r=\arg\min_{1\leq j\leq R} \left\{\frac{\wh\lambda_{j+1}}{\wh\lambda_j}\right\},
\end{equation}
where we choose $R=p/2$ as suggested in their paper. Figure~A1  of the online 
supplement presents 
the boxplots of $\wh r$.  We see from 
Figure~A1 that the estimated number of factors $\wh r$ is the sum of the 
number of common factors and the number of spiked components of the noises. 
The result indicates that 
the ratio-based method may fail to identify the correct number of factors with dynamic dependencies 
if the covariance matrix of the noise has diverging eigenvalues. 
On the other hand,  the high-dimensional white noise test considered in the paper 
continues to work well.


\begin{table}
 \caption{Empirical probabilities $P(\hat{r}=r)$ for Example 2 with $r=5$, $K=3$ or 7, where $p$ and $n$ are the dimension and the sample size, respectively. $\delta_1$ and $\delta_2$ are the strength parameters of the factors and the errors, respectively. $1000$ iterations are used.}
          \label{Table2}
\begin{center}
 \setlength{\abovecaptionskip}{0pt}
\setlength{\belowcaptionskip}{3pt}

\begin{tabular}{c|c|ccccc|ccccc}
\hline
 & &\multicolumn{5}{c}{$n$ ($K=3$)}&\multicolumn{5}{c}{$n$ ($K=7$)}\\
$(\delta_1, \delta_2)$ &$p$&$300$&$500$&$1000$&$1500$&$3000$&$300$&$500$&$1000$&$1500$&$3000$\\
\hline
 (0,0) &$50$&0.510&0.833&0.906&0.917&0.926&0.418&0.688&0.904&0.908&0.910\\
 &$100$&0.538&0.799&0.910&0.916&0.922&0.426&0.754&0.910&0.916&0.918\\
&$300$&0.582&0.907&0.916&0.924&0.932&0.406&0.686&0.914&0.925&0.926\\
&$500$&0.560&0.888&0.918&0.928&0.932&0.614&0.778&0.912&0.918&0.920\\
\hline
(0.4,0.5) &$50$&0.717&0.903&0.928&0.929&0.935&0.806&0.820&0.892&0.912&0.926\\
&$100$&0.800&0.924&0.938&0.940&0.944&0.800&0.914&0.922&0.904&0.922\\
&$300$&0.858&0.904&0.928&0.932&0.952&0.939&0.935&0.935&0.929&0.930\\
&$500$&0.834&0.922&0.932&0.933&0.948&0.898&0.904&0.926&0.930&0.933\\
   \hline 
   (0.5,0.4) &$50$&0.420&0.890&0.910&0.916&0.920&0.332&0.856&0.900&0.928&0.938\\
 &$100$&0.508&0.868&0.912&0.928&0.936&0.356&0.716&0.920&0.922&0.928\\
 &$300$&0.581&0.910&0.926&0.929&0.932&0.384&0.688&0.924&0.936&0.945\\
 &$500$&0.678&0.928&0.936&0.938&0.934&0.421&0.778&0.924&0.930&0.931\\
\hline
\end{tabular}
          \end{center}
\end{table}

\begin{figure}
\begin{center}
{\includegraphics[width=0.6\textwidth]{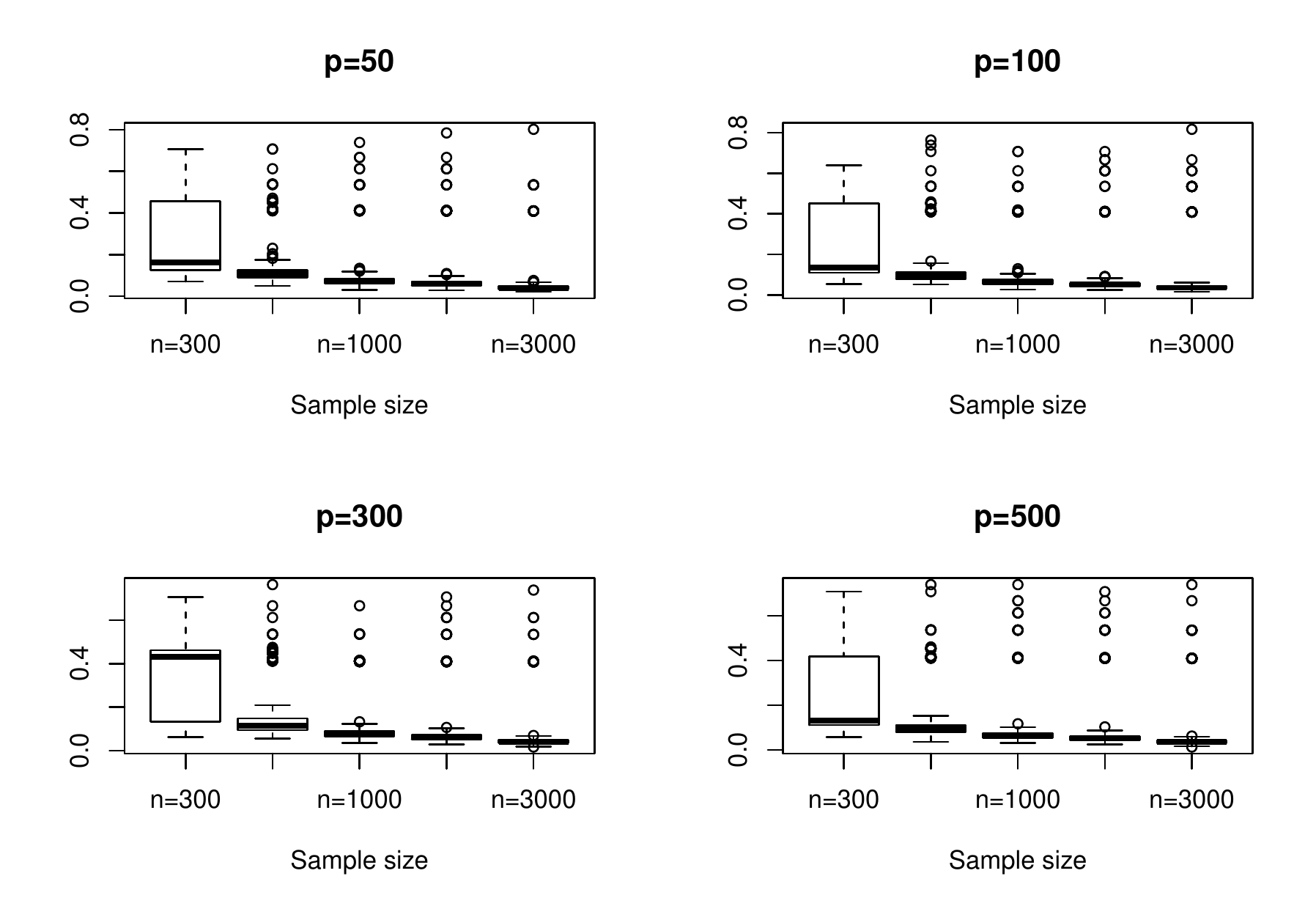}}
\caption{Boxplots of $\bar{D}(\wh\bA_1,\bL_1)$ when $r=5$ and $K=3$ under the scenario that $p$ is relatively large in Example 2. The sample sizes used are $n=300,500,1000,1500,3000$ 
and the number of iterations is $1000$.}\label{fig5}
\end{center}
\end{figure}

Next, we study the accuracy of the estimated loading matrices as that in Example 1. 
The boxplots of $\bar{D}(\wh\bA_1,\bL_1)$ are shown in Figure~\ref{fig5}. Similar patterns are also obtained  for the estimation of other matrices, 
and we omit them here. From Figure~\ref{fig5}, there is a clear pattern that the estimation 
accuracy of the loading matrix improves as the sample size increases even for moderately large $p$, 
which is in line with our asymptotic theory. The results also confirm that the proposed white noise 
test selects $\hat{r}$ reasonably well even for large $p$.

Finally, since a key difference between the proposed method and that of LYB is that 
we allow some of the eigenvalues of the noise covariance to diverge, it is of interest to 
compare the two methodologies. 
Denoting the proposed method by GT, we compared it with LYB 
 in terms of the RMSE defined below:
\begin{equation}\label{rmse:plg}
\text{RMSE}=\left(\frac{1}{np}\sum_{t=1}^n\|\wh\bA_1\wh\bx_t-\bL_1\bff_t\|_2^2\right)^{1/2},
\end{equation}
which is different from that in Equation (\ref{rmse:psm}) because we have another 
factor $p^{-1/2}$ in (\ref{rmse:plg}). This RMSE quantifies the estimation accuracy of the 
common factor process. In the comparison, the number of factors is obtained by the 
corresponding method of each methodology. 
The results are shown in Table~\ref{Table4}  for $r=5$, $K=7$ and $\delta_1=\delta_2=0$. The pattern is similar for the other settings of $\delta_1$ and $\delta_2$.    When calculating $\wh{\bB}_2^*$ using our method, we choose the number of components $\wh K=10$, which is fixed for all the iterations. Thus, $\wh{\bB}_2^*$ contains $p-\wh K$ columns corresponding to the $p-\wh K$ smallest eigenvalues of $\wh \bS$. From the table, we see that, because the ratio-based method tends to overestimate the 
number of common factors $r$ in the presence of diverging eigenvalues in the covariance 
matrix of the idiosyncratic component, 
the RMSE of our method is much smaller than that obtained by \cite{LamYaoBathia_Biometrika_2011}. Also, as expected, for a given $p$, the RMSE tends to decrease when the sample size increases. This is in agreement with the asymptotic theory in Theorem \ref{tm4}. In the online supplement, we fix the number of factors and compare our method with that of LYB using the RMSE criterion to check the need to mitigate  the noise effect. 
Our method continues to produce smaller RMSEs in general when the noise effect is diverging. 
See Table A8 of the supplement for details. Overall, under the 
assumption that the top eigenvalues of the noise covariance matrix are diverging in 
high-dimensional time series, 
the proposed method outperforms the existing one in the literature.

\begin{table}
 \caption{The RMSE defined in (\ref{rmse:plg}) when $r=5$ and $K=7$ in Example 2.  
 The sample sizes used are $n=300, 500, 1000, 1500, 3000$. Standard errors are given in the parentheses and $1000$ iterations are used. GT denotes the proposed method and ‘LYB’ is the one in \cite{LamYaoBathia_Biometrika_2011}}
          \label{Table4}
          
\footnotesize{
\begin{center}
 \setlength{\abovecaptionskip}{0pt}
\setlength{\belowcaptionskip}{3pt}

\begin{tabular}{c|c|ccccc}
\hline
&&\multicolumn{5}{c}{$n$}\\
Method&$p$&$300$&$500$&$1000$&$1500$&$3000$\\
\hline
 GT&$50$& 1.510(0.233)& 1.124(0.235)& 0.770(0.235)& 0.627(0.224)& 0.488(0.273)\\
LYB&&3.056(0.085)&3.051(0.081)&3.056(0.075)&3.053(0.122)&2.976(0.400)\\
\hline
 GT&$100$& 1.490(0.179)& 1.148(0.188)& 0.817(0.141)& 0.677(0.126)& 0.519(0.191)\\
LYB&&3.050(0.074)&3.056(0.065)&3.053(0.055)&3.046(0.159)&3.024(0.257)\\
\hline
 GT&$300$& 1.729(0.118)& 1.463(0.107)& 1.149(0.094)& 1.107(0.079)& 0.769(0.077)\\
LYB&&3.052(0.047)&3.055(0.047)&3.053(0.040)&3.056(0.037)&3.056(0.034)\\
\hline
 GT&$500$& 1.753(0.089)& 1.547(0.081)& 1.285(0.052)& 1.044(0.070)& 0.861(0.047)\\
LYB&&3.057(0.053)&3.050(0.042)&3.054(0.035)&3.055(0.034)&3.055(0.027)\\
\hline
\end{tabular}
          \end{center}}
\end{table}


\subsection{Real data analysis}
In this section, we demonstrate the application of the proposed method using 
U.S. monthly macroeconomic data with 108 time series and 283 observations.  
An additional real example is shown in the online supplement, where 
the dimension $p$ is greater than the sample size $n$. 

{\noindent \bf Example 3.}  In this example, we employ the data set of 
monthly U.S. macroeconomic time series, which is available from \url{https://www.princeton.edu/~mwatson/publi.html} and used in 
\cite{StockWatson_2009}. There are 108 variables available in the data file 
spanning from 1959 to 2006. 
To remove any possible trend in the series, 
we take the first difference $\by_t=\wt\by_t-\wt\by_{t-1}$, where $\wt\by_t\in R^{108}$ and $t=2,...,576$. The differenced series are shown in Figure~A2 of the online supplement, from which we see that the series continue to display some strong nonstationary patterns between later 1990s and early 2000s. This might be due to several economic and financial crises in the US and around the world in the period, including the 1997 Asian financial crisis and the 9/11 attack in 2001. Therefore, we  use the data from December 1972 to July 1996 only.  
The resulting differenced U.S. monthly macroeconomic time series 
are shown in Figure~A3  of the online supplement with $n=283$ and $p=108$.

Following the proposed modeling procedure, 
we first applied the proposed white-noise tests of Section 2.3 to identify the number of common factors and found $\wh r=8$. 
We also calculated the eigenvalues of $\wh\bS$, and Figure~A4(a) of the 
online supplement plots the first ten eigenvalues. We see clearly that, in this 
particular instance, the largest eigenvalues are extremely large because the average of the first ten eigenvalues of $\wh\bSigma_y$ is $1.0\times 10^6$. In our analysis, 
we choose $\wh K=5$, but the results are similar for other choices. The spectral densities of the eight estimated common factors are given in Figure~A5 of the supplement, and they show clearly that the estimated factors are all serially dependent processes. In addition, the variance of the first factor is extremely large compared to those of the others. We also applied the method of LYB to $\by_t$, and the estimated number of factors is $\wh r=1$.  We further present the spectral densities of the first six transformed series $\wh u_{1t},...,\wh u_{6t}$ using the eigen-analysis of LYB in Figure~A6 of the supplement and find that  none of  these series is white noise.  
This marks a contradiction with the assumptions used in LYB, which assume that $\wh u_{2t}$ to $\wh u_{6t}$ are white noises. 

Next we examine the forecasting performance of the extracted factors via different methods. We estimate the models using the data in the time span $[1,\tau]$ with $\tau=183,...,283-h$ for the $h$-step ahead forecasts. First, we study the overall performance by employing a VAR(1) or an AR(1) model for the estimated factor processes and compute $h$-step ahead predictions of $\by_t$ using the associated factor loadings. The forecast error is defined as 
\begin{equation}\label{fe}
\text{FE}_h=\frac{1}{100-h+1}\sum_{\tau=183}^{283-h}E(\tau,h) \quad \mbox{with}\quad 
E(\tau,h) = \frac{1}{\sqrt{p}}\|\wh\by_{\tau+h}-\by_{\tau+h}\|_2,
\end{equation}
where $p=108$. Table~\ref{Table41} reports the 1-step to 3-step ahead forecast errors using VAR(1) or AR(1) models, where we also used the first 8 factors obtained by the method in \cite{BaiNg_Econometrica_2002} although it identifies 30 factors. The smallest forecast error of  each step is shown in boldface. From the table, we see that, except for the 1-step ahead forecast, our method is capable of producing accurate forecasts and the associated forecast errors based on the extracted factors by our method are smaller than those based on the other two methods. The PCA method 
also works well in the comparison. 
Next, we further examine the forecastability of the extracted factors by different methods. We adopt the diffusion index model as in \cite{StockWatson_2002a} and \cite{StockWatson_2002b} and predict  some particular components of $\by_t$. We focus on two important components: (a) CPI-All: 
the Consumer Price Index for All Urban Consumers: All Items in U.S. City Average, and  
(b) CPI-Core: the Consumer Price Index for All Urban Consumers: All Items Less Food and Energy in U.S. City Average. These two components are often used to measure the  inflation.  The difference between CPI-All and CPI-Core is that the latter does not include the more volatile categories of food and energy prices. To see the forecastability of the extracted factors, we only regress $y_{i,t+h}$ on $\wh \bff_t$ without including the lagged variable of $y_{it}$, 
 because  \cite{StockWatson_2002b} found that the simpler diffusion index forecasts excluding the lags are even better than that with the lagged ones. Table \ref{Table42} presents the 1-step to 3-step ahead mean squared forecast errors for the CPI-All and the CPI-Core, where F-DI denotes the factor-based diffusion index model as in \cite{StockWatson_2002b} without the lagged variables, and F-AR represents the factor-based VAR or AR approach mentioned before. 
From Table \ref{Table42}, we see that, for factor-based diffusion index models, the forecastability of the factors extracted by our method  fares better than those of the other two methods, which is understandable since our 
extracted factors capture most of the dynamic dependencies. For the factor-based VAR or AR approach, our method is not as good as the PCA approach, but better than the LYB, in forecasting the CPI-All, but it fares better than the other two approaches 
in predicting the CPI-Core. In conjunction with the results of Table \ref{Table41}, our approach fares better in terms of the overall MSE defined in (\ref{fe}), but it may not perform uniformly better than the others for every component of $\by_t$.

\begin{table}[h]
 \caption{The 1-step to 3-step ahead forecast errors. GT denotes the proposed method, ‘BN’ denotes the principal component analysis as in \cite{BaiNg_Econometrica_2002}, and ‘LYB’ is the one in \cite{LamYaoBathia_Biometrika_2011}. 
 Boldface numbers denote the smallest one for a given model.}
          \label{Table41}
\begin{center}
 \setlength{\abovecaptionskip}{0pt}
\setlength{\belowcaptionskip}{3pt}

\begin{tabular}{c|ccccc}
\hline
&&&\multicolumn{3}{c}{Methods}\\
\cline{4-6}
Step&Model&&GT&BN&LYB\\
\hline
1 & AR(1) && 264.1 & {\bf 260.1} & 347.7 \\
2&AR(1)&&{\bf 286.1}&288.1&359.4\\
3&AR(1)&&{\bf 306.1}&309.7&379.4\\
\hline
\end{tabular}
          \end{center}
\end{table}

\begin{table}
 \caption{The 1-step to 3-step ahead mean squared forecast errors for inflation 
 indexes. GT denotes the 
 proposed  method, ‘BN’ denotes the principal component analysis as in \cite{BaiNg_Econometrica_2002} and ‘LYB’ is the one in \cite{LamYaoBathia_Biometrika_2011}. 
 Boldface numbers denote the smallest one for a given model. CPI - All denotes 
 the Consumer Price Index for All Urban Consumers: All Items in U.S. City Average, and  CPI - Core denotes the Consumer Price Index for All Urban Consumers: All Items Less Food and Energy in U.S. City Average. F-DI denoted factor-based diffusion index model and 
 F-AR denotes factor-based VAR or AR models.}
          \label{Table42}
\begin{center}
 \setlength{\abovecaptionskip}{0pt}
\setlength{\belowcaptionskip}{3pt}

\begin{tabular}{c|cccccccc}
\hline
&&\multicolumn{3}{c}{CPI - All}&&\multicolumn{3}{c}{CPI - Core}\\
\cline{3-5}\cline{7-9}\\
Step&Model&GT&BN&LYB&&GT&BN&LYB\\
\hline
1 & F-DI & {\bf 0.126} & 0.209 & 0.203 & &{\bf 0.111} & 0.216 & 0.212 \\
   & F-AR & 0.247 & {\bf 0.218} & 0.229 & &{\bf 0.159} & 0.160 & 0.225 \\ 
   \hline 
2&F-DI&{\bf 0.128}&0.206&0.188&&{\bf 0.113}&0.214&0.218\\
&F-AR&0.244&{\bf 0.217}&0.228&&{\bf 0.164}&0.165&0.224\\
\hline
3&F-DI&{\bf 0.135}&0.207&0.191&&{\bf 0.113}&0.215&0.224\\
&F-AR&0.243&{\bf 0.219}&0.229&&{\bf 0.167}&0.168&0.224\\
\hline
\end{tabular}
          \end{center}
\end{table}
 
To further compare the forecast ability of extracted factors by different approaches, 
we employ the diffusion index models and adopt the asymptotic  test in \cite{diebold-1995} to test for equal predictive ability.  The null hypothesis of interest is that the approaches considered have equal predictive ability, and the alternative is that our proposed method performs better than the others in predicting the inflation indexes.
Table~\ref{Table43} reports the testing results, where we use the method in \cite{andrews1991} to calculate the long-run covariance matrix (L-COV). In the table, GT--BN and GT--LYB denote the comparison between our approach and that in BN and LYB, respectively, and the test statistic is the difference between the forecast errors of the two methods involved, as illustrated in Diebold and Mariano (1995). From Table~\ref{Table43}, we see that all $p$-values of the test statistics are 
small with most of them being less than or equal to 0.05, indicating that the factors 
extracted by our method have similar or better forecast ability than those of the other two 
methods  in predicting the CPI indexes. 

In conclusion, for U.S. monthly macroeconomic variables, the forecasting performance of the extracted factors using diffusion index models seems to favor the proposed approach, since the factors extracted by our method 
capture most of the dynamic dependencies of the data whereas 
those by the other methods may overlook some dynamic dependence. 
On the other hand, for the forecasting performance of the factor-based VAR or AR models, our method only fares better for the CPI-Core and it is not as good 
as the PCA approach for the CPI-All. 

Finally, we state that the proposed factor model is different from those available in the literature such as those considered in \cite{BaiNg_Econometrica_2002}. Therefore, the factors extracted by the proposed method are different from those by PCA and they have different interpretations. We believe that the PCA approach and the proposed one are complementary to each other rather than competitors, because they explore different aspects of the data as illustrated in the data  analysis. The proposed method can be treated as another option in the toolbox for modeling high-dimensional time series 
and the extracted factors can be useful to practitioners who are interested in 
out-of-sample forecasting. 

\begin{table}
\caption{Testing for equal predictive ability of different factor-based diffusion index models, using the asymptotic test of \cite{diebold-1995}. GT denotes the proposed method, ‘BN’ denotes the principal component analysis  as in \cite{BaiNg_Econometrica_2002}, and ‘LYB’ is the one in \cite{LamYaoBathia_Biometrika_2011}. 
CPI - All denotes the Consumer Price Index for All Urban Consumers: All Items in U.S. City Average, and  CPI - Core denotes the Consumer Price Index for All Urban Consumers: All Items Less Food and Energy in U.S. City Average.} 
          \label{Table43}
\begin{center}
 \setlength{\abovecaptionskip}{0pt}
\setlength{\belowcaptionskip}{3pt}

\begin{tabular}{c|cccccc}
\hline
&&\multicolumn{2}{c}{CPI - All}&&\multicolumn{2}{c}{CPI - Core}\\
\cline{3-4}\cline{6-7}\\
Step&&GT--BN&GT--LYB&&GT--BN&GT--LYB\\
\hline
1 & L-COV&0.0026  & 0.0025 & &0.007&  0.0023 \\
   & $p$-value& $0.05$&  $0.06$& &$0.10$  &$0.02$ \\ 
   \hline 
2&L-COV&0.0022&0.0025&&0.001&0.002\\
&$p$-value&$0.03$&$0.11$&&$\approx 0$&$0.01$\\
\hline
3&L-COV&0.002&0.002&&0.001&0.002\\
&$p$-value&$0.05$&$0.10$&&$\approx 0$&$0.01$\\
\hline 
\end{tabular}
          \end{center}
\end{table}

\section{Discussion and Concluding Remarks}
This article introduced an alternative factor model for 
high-dimensional time series analysis. 
We allow the largest eigenvalues of the covariance matrix of the idiosyncratic components 
to diverge to infinity by imposing some structure on the noise terms. 
The first step of the proposed analysis 
is an eigen-analysis of the matrix $\wh\bM$ defined in Equation 
(\ref{mhat}). The form of $\wh\bM$ is a special case of the orthonormalized Partial Least Squares 
for  time series data by assuming the covariance matrix of the data is identity. By an abuse of notation, let $\bw_t=(\by_{t-1}',...,\by_{t-k_0}')'$ be the vector of past $k_0$ lagged values of 
the time series $\by_t$, 
where $k_0$ is a pre-specified positive integer as that in (\ref{M}) and (\ref{mhat}). The orthonormalized Partial Least Squares computes the orthogonal score vectors for $\by_t$ by solving the following optimization problem:
\begin{equation}\label{pls:sp}
\max_{\ba_i}\|E(\ba_i'\by_t\bw_t')\|_2^2,\quad \text{subject to}\quad \ba_i' E(\by_t\by_t')\ba_i=1.
\end{equation}
See, for example, \cite{arenas2008}. It can be shown that the columns $\ba_i$ are given
by the  principal eigenvectors of the following generalized
eigenvalue problem:
\begin{equation}\label{gn:ev}
\bSigma_{yw}\bSigma_{yw}'\ba_i=\eta\bSigma_y\ba_i.
\end{equation}
Note that $\bM=\bSigma_{yw}\bSigma_{yw}'$, which is just the form in (\ref{M}). 
To solve the above equation, we need to obtain accurate estimates for the covariance matrix 
and its inverse simultaneously, which however is not easy. 
 Instead we change the subject condition in (\ref{pls:sp}) to $\ba_i'\ba_i=1$ and  apply the eigen-analysis on  $\wh\bM$ in (\ref{mhat}), and this approach 
 remains an effective way if we assume the component variances of the data are 
 uniformly bounded. In this case, the second step is needed.

The second step of the proposed analysis is the projected PCA on $\wh\bS$ in (\ref{cs}) by assuming the largest $K$ eigenvalues of the covariance matrix of the idiosyncratic component are diverging. 
In practice, one may assume that the largest eigenvalue is diverging whereas the  
remaining ones are bounded. Limited experience indicates that many real datasets 
exhibit such a phenomenon. 
If we are only concerned with the forecasting performance of the proposed analysis, 
we may select $\wh K$ in a range such as $\wh K_L\leq \wh K\leq \wh K_U$, where $\wh K_L$ and $\wh K_U$ are defined in Remark 5(ii), via some cross-validation method like the out-of-sample 
testing. 

The white noise test employed is an efficient way to determine the number of common factors for the proposed factor model. 
The one-by-one bottom-up testing procedure may not perform well when the 
dimension $p$ is large. On the other hand, the limiting distribution of the test statistic of \cite{Tsay_2018} holds for large $p$ by making use of the extreme value theory, and, hence, the test fares well for large $p$. If we like to use the test  statistic 
for a wide range of dimensions and various sample sizes, 
we may employ empirical critical values via some simulation, which is possible because 
the test uses rank correlations and is robust to the underlying distribution of the data under the null hypothesis. The simulation results of \cite{Tsay_2018} show that the resulting test statistic works reasonably well.

\section*{Supplementary Material}
The supplementary material contains all  technical proofs of the theorems in Section 3 and an additional real example consisting of half-hourly temperature data observed at the 
Adelaide Airport in Australia with $p$ = 508 and $n$ = 336. 

\section*{Acknowledgments}
 We are grateful to the  Editor, Associate Editor  and the anonymous referees for their insightful comments and suggestions that have substantially improved the presentation and quality 
 of the paper.  This research is supported in part by the Booth School of Business, University of Chicago.


\end{document}